\documentclass[a4paper,11pt]{article}
\usepackage{jheppub} 
\usepackage{slashed}
\usepackage{bbm,mathtools}
\usepackage{multirow,booktabs}
\usepackage{colortbl}
\usepackage{subcaption}
\usepackage{array,arydshln}
\usepackage[usenames,dvipsnames,svgnames,table]{xcolor}
\usepackage{enumitem}
\usepackage{makecell}

\usepackage{siunitx}
\usepackage{placeins}
\definecolor{Gray}{gray}{0.95}

\usepackage{dsfont}

\usepackage{xparse}
\ExplSyntaxOn
\NewDocumentCommand{\longdash}{ O{2} }
{
	--\prg_replicate:nn { #1 - 1 } { \negthinspace -- }
}
\ExplSyntaxOff

\definecolor{nicered}{rgb}{0.6,0.1,0.1}
\definecolor{nicegreen}{rgb}{0.1,0.5,0.1}
\definecolor{mediumcandyapplered}{rgb}{0.99, 0.12, 0.07}
\definecolor{red}{rgb}{1.0, 0, 0}
\hypersetup{colorlinks,citecolor= nicegreen,linkcolor= nicered}

\def\eg{\hbox{\it e.g.}{}}

\newcommand{\U}{\mathbf{U}}
\newcommand{\V}{\mathbf{V}}

\newcommand{\Lag}{\mathcal{L}}
\newcommand{\F}{\mathcal{F}}
\newcommand{\de}{\partial}

\newcommand{\meLabel}{\mathcal{M}}

\title{
Higgs pair production in gluon fusion to higher orders in Higgs Effective Field Theory 
}

\author[a,b]{Ilaria Brivio,}
\author[c,d]{Ramona Gr\"{o}ber,}
\author[c,d]{Konstantin Schmid}

\emailAdd{ilaria.brivio@unibo.it}
\emailAdd{ramona.groeber@pd.infn.it}
\emailAdd{konstantin.schmid@pd.infn.it}

\affiliation[a]{Dipartimento di Fisica e Astronomia, Universit\`a di Bologna,  I-40126 Bologna, Italy}
\affiliation[b]{Istituto Nazionale di Fisica Nucleare, Sezione di Bologna, I-40126 Bologna, Italy}
\affiliation[c]{Dipartimento di Fisica e Astronomia ``G. Galilei", Universit\`a di Padova, Sezione di Padova, I-35131 Padova, Italy}
\affiliation[d]{Istituto Nazionale di Fisica Nucleare, Sezione di Padova, I-35131 Padova, Italy}

\abstract{
Higgs pair production offers the opportunity to probe correlations among the couplings of one or two Higgs bosons to fermions and gauge bosons. In this context, it serves as a powerful test of the underlying Effective Field Theory (EFT) framework. In particular, while such couplings remain correlated in the Standard Model Effective Field Theory (SMEFT) at dimension six, they can become fully de-correlated in Higgs Effective Field Theory (HEFT) already at leading order in the EFT expansion.
In this work, we study Higgs pair production via gluon fusion within the HEFT framework. We demonstrate that adopting a consistent power counting in combination with next-to-leading order (NLO) diagrams necessitates the inclusion of higher-dimensional operators beyond the leading ones. We analyze their phenomenological impact and re-assess critically the kinematic benchmark scenarios commonly used in experimental non-resonant di-Higgs searches in light of these additional contributions.}

\date{}

\subheader{\hfill\rm COMETA-2025-52}

\begin{document} 

\maketitle

\section{Introduction}
An ultimate probe of the Standard Model (SM) Higgs sector is provided by the measurement of multi-Higgs production, allowing to access interactions with more than a single Higgs boson involved. The measurement of multi-Higgs production is very challenging due to small cross sections: the Higgs pair production cross section is roughly speaking three orders of magnitude lower than single Higgs production and is hence still rather weakly bound. 
Nevertheless, projections for the HL-LHC show that sensitivity to the SM cross section can be reached \cite{CMS:2022dwd, ATL-PHYS-PUB-2025-001, CMS:2025hfp}.
From an experimental perspective, this has become possible due to improvements in analysis methods such as $\tau$-reconstruction and $b$-tagging. These advancements are crucial for the identification of final states like $b\bar{b}b\bar{b}$, $b\bar{b} \tau^+ \tau^-$, and $b\bar{b} \gamma \gamma$ \cite{ATLAS:2022faz, CMS:2022dwd}, which are the most promising ones for a discovery of a Higgs boson pair.

The measurement of Higgs pair production allows to probe various beyond-the-Standard Model (BSM) scenarios. In particular, it provides a direct probe of the trilinear Higgs self-coupling \cite{Djouadi:1999rca, Dolan:2012rv, Baglio:2012np}, whose measurement has direct implications for electroweak baryogenesis \cite{Curtin:2014jma} or Higgs portal models which can arise in the context of dark matter or neutral naturalness models \cite{Haisch:2022rkm, Haisch:2023aiz}.
In the SM, the dominant process is gluon fusion \cite{DiMicco:2019ngk} where apart from modifications of the trilinear Higgs self-coupling, various other new physics effects could modify the cross section, such as for instance new particles running in the loop as they arise in the minimal supersymmetric extension of the SM (MSSM) \cite{Plehn:1996wb, Agostini:2016vze} or in composite Higgs models \cite{Gillioz:2012se, Grober:2016wmf, DeCurtis:2023pus}. 

In the recent years, a bottom-up approach to new physics  in terms of effective field theories (EFTs) has become quite established. There are two candidate EFTs: Standard Model Effective Field Theory (SMEFT), where the leading BSM effects in Higgs physics stem from dimension-six operators \cite{Grzadkowski:2010es} and Higgs Effective Field Theory (HEFT) \cite{Feruglio:1992wf, Burgess:1999ha, Azatov:2012bz, Buchalla:2013rka, Alonso:2012px, Brivio:2013pma}, which originates from chiral perturbation theory. 

In this paper, we will study Higgs pair production in gluon fusion in the more general EFT -- HEFT.
This is motivated since Higgs pair production provides a probe whether electroweak symmetry breaking is realized linearly or non-linearly.\footnote{See e.g. \cite{Grober:2010yv, Contino:2012xk} for early works on Higgs pair production as a probe of the non-linear character of a theory.} Higgs pair production in gluon fusion has been computed up to NLO QCD in \cite{Buchalla:2018yce, Heinrich:2020ckp}\footnote{For computations in the infinite top mass limit, see the earlier work \cite{Grober:2015cwa} and \cite{Grober:2017gut} for CP-violating operators. See \cite{deFlorian:2021azd} for a computation with anomalous couplings at NNLO QCD employing the infinite top mass limit for the NNLO amplitudes. In the SM, the cross section is known up to $\text{N}^3$LO QCD for infinite top mass \cite{Chen:2019fhs, Chen:2019lzz, Chen:2026zmi} and including $\text{N}^3$LL resummation~\cite{AH:2022elh}. Ref.~\cite{Grazzini:2018bsd} provides NNLO QCD corrections with approximate top quark mass effects.} using the LO HEFT Lagrangian building on the NLO QCD corrections in full top quark mass dependence as presented in Refs.~\cite{Borowka:2016ehy, Borowka:2016ypz} and have been computed with alternative methods in Refs.~\cite{Baglio:2020ini, Bagnaschi:2023rbx, Davies:2025qjr}.
The impact of HEFT operators entering in one-loop corrections to the Higgs self-interaction have been discussed in \cite{Anisha:2024ljc}. While we will also compare our results to SMEFT at the strictly linear level, we refer to more detailed studies of the effects of dimension-six operators in the SMEFT in Higgs pair production to Refs.~\cite{Heinrich:2022idm, Alasfar:2023xpc, Heinrich:2023rsd, DiNoi:2024ajj,  Heinrich:2024rtg, Maltoni:2024dpn}. For a study of the impact of dimension-8 operators we refer to future work \cite{ggHHdim8}. Higgs pair production has been investigated with on-shell amplitude methods as well~\cite{Shadmi:2018xan,Liu:2023jbq,Grober:2025vse},  see in particular Ref.~\cite{Grober:2025vse} for the matching of SMEFT and HEFT Higgs-gluon operators.
\par
Our study differs by the ones presented in literature by the addition of next-to-leading order (NLO) and next-to-next-to-leading order (NNLO) operators in the HEFT expansion, which should be considered when employing a consistent power counting of the HEFT expansion as proposed by us in \cite{Brivio:2025yrr}. At the same time, the study presented here can be considered as a prime example of the power counting method of \cite{Brivio:2025yrr}: due to the simplicity of the process, containing only gluons, Higgs bosons and top quarks, the process allows to consider the effects of the higher-order operators. Furthermore, it is a highly non-trivial study case as the process is loop-induced in the SM, while in EFTs it can arise also at tree-level. 
\par
We study in detail the phenomenological implications of the operators with higher chiral dimension. In particular, we revisit the kinematic benchmarks proposed in \cite{Carvalho:2015ttv,Capozi:2019xsi, Alasfar:2023xpc} and connect the number of distinguishable benchmark scenarios to experimental and theoretical uncertainties on the invariant Higgs mass distribution. While the current coverage is very good, rather rare parameter choices in HEFT can still produce distributions outside the existing categories.
\par
On a similar note, Ref.~\cite{Englert:2025xrc} considers kinematic corrections to the trilinear Higgs self-coupling in HEFT. Our study is more general, including further operators.
The paper is structured as follows: in Section~\ref{sec_powercounting_recap} we review HEFT and its power counting, in Section \ref{sec_Di_Higgs_prod} we show our calculation of the HEFT Higgs pair production cross section and perform the phenomenological analysis. We conclude in Section \ref{sec_conclusions}.

\section{Higgs Effective Field Theory and power counting prescriptions} 
\label{sec_powercounting_recap}
In this section we will introduce the EFT descriptions used in this work. We mostly discuss HEFT, sometimes also referred to as electroweak chiral Lagrangian, for which we will introduce our notation in this section. We will though also compare our findings with SMEFT at dimension-six level, for which we adopt the so-called Warsaw basis \cite{Grzadkowski:2010es}. For our SMEFT notation we refer directly to Subsection~\ref{sec:SMEFT}, where we present our SMEFT results.
\subsection{The HEFT Lagrangian}
While in the SM (and in SMEFT) the Higgs boson is constrained to reside within an $SU(2)_L$ doublet, HEFT relaxes this requirement, allowing for a more general description of its dynamics.
 The three Goldstone bosons (GBs) responsible for the correct infra-red (IR) behaviour of the gauge sector arise independently of the physical Higgs boson, and are incorporated into the complex matrix
\begin{align}
    \U = \exp \left( \frac{i\,\pi_I \sigma^I}{v}\right)\,,
\end{align}
where $\pi^I$ are the GBs in a nonlinear representation, and $\sigma^I$ are the Pauli matrices. As a consequence the HEFT description is limited to scales $\Lambda < 4 \pi v$.\footnote{This can be relaxed by introducing another scale $f$, as e.g. in Composite Higgs Models \cite{Contino:2006qr}. The introduction of such a scale can assimilate HEFT with SMEFT \cite{Giudice:2007fh} but -- in a pure bottom-up context -- is somewhat ambiguous.
}
\par
The matrix $\U$~transforms as a bidoublet under the global $SU(2)_L \times SU(2)_R$ symmetry with its covariant derivative given by
\begin{align}
    D_{\mu} \U = \partial_{\mu} \U + \frac{i g}{2} W_{\mu}^I\, \sigma^I \U - \frac{i g'}{2} B_{\mu}\, \U \sigma^3\,.
\end{align}

Since the physical Higgs boson $h$ is introduced as a gauge singlet, the $h$ field can be arbitrarily inserted in any operator.
Furthermore, as will be discussed later, its insertion does not increase the HEFT power counting, so that any polynomial in the Higgs field contributes at the same order. This implies that power series of  $h$ insertions up to arbitrarily large powers provide the building blocks of the Lagrangian, for which one can define the dimensionless functionals~\cite{Grinstein:2007iv}
\begin{align}
\label{eq.general_flare_function}
    \F_i(h) &= d_{i} + a_i \frac{h}{v} + b_i \frac{h^2}{v^2} + \dots, 
    &
    d_i, a_i, b_i, \dots &\in \mathbb{C}\,.
\end{align}
The coefficients are complex in general, but, if $\F_i$ is inserted in a hermitian operator, only their real parts are physical. In the following, we will adopt the notation that coefficients with $d$ denote the first element in the series, coefficients denoted with $a$ the second element associated with one power in the physical Higgs $h$, etc.

The full HEFT Lagrangian is an expansion in higher dimensional operators whose criteria of expansion will be reviewed in a bit. In the meanwhile we will write
\begin{equation}
    \Lag_{\rm HEFT} = \Lag_{\rm LO} + \Lag_{\rm NLO} + \Lag_{\rm NNLO}+\dots\,,
\end{equation}
where the LO term is conventionally taken to be \cite{Appelquist:1980vg, Longhitano:1980tm, Feruglio:1992wf, Appelquist:1993ka}:
\begin{align}
    \Lag_{\rm LO} = &
    - \frac{1}{4} G_{\mu \nu}^a G^{a \mu \nu} - \frac{1}{4} W^I_{\mu \nu} W^{I \mu \nu} - \frac{1}{4} B_{\mu \nu} B^{\mu \nu} 
    + \frac{1}{2} \partial_{\mu} h \partial^{\mu} h - \frac{v^2}{4} \mathrm{Tr}\left(\V_{\mu} \V^{\mu}\right) \mathcal{F}_C(h) - \lambda v^4\mathcal{V}(h) 
    \nonumber\\
    &+ i \overline{Q}_L \slashed{D} Q_L + i \overline{Q}_R \slashed{D} Q_R + i \overline{L}_L \slashed{D} L_L + i \overline{L}_R \slashed{D} L_R 
    \nonumber\\
    &- \frac{v}{\sqrt{2}} \left(\overline{Q}_L \U \mathcal{Y}_Q(h) Q_R + \mathrm{h.c.} \right) - \frac{v}{\sqrt{2}} \left(\overline{L}_L \U \mathcal{Y}_L(h) L_R + \mathrm{h.c.} \right).
\label{eq:lagrangian_HEFT_LO}
\end{align}
Here, we did not insert any custodial symmetry breaking or CP-violating operator and wrote the fermion fields in such a way to keep the global $SU(2)_L\times SU(2)_R$ manifest
\begin{align}
    Q_L &= (u_L, d_L)\,, & Q_R &= (u_R, d_R)\,, & L_L &= (\nu_L, e_L)\,, & L_R =& (0, e_R)\,.
\end{align}
We have introduced in the object
\begin{align}
 \V_{\mu} &= (D_{\mu} \U) \U^{\dagger}\,,
\label{eq:defV}
\end{align}
into Eq.~\eqref{eq:lagrangian_HEFT_LO}, which transforms in the adjoint representation of $SU(2)_L$: $\V_{\mu} \rightarrow \Omega_L \V_{\mu} \Omega_L^{\dagger}$. The dimensional functionals concretely become
\begin{align}
    \F_C(h) &= 1 + a_C \frac{h}{v}+b_C\frac{h^2}{v^2} +...\,,
    \\
    \mathcal{V}(h) &= \frac{h^2}{v^2} + a_{\lambda^3} \frac{h^3}{v^3} + b_{\lambda^3}  \frac{h^4}{4v^4} + ...
    \\
    \mathcal{Y}_Q(h) &= {\rm diag}\left(\mathcal{Y}_U(h),\, \mathcal{Y}_{D}(h)\right)\,,
    \qquad
    \mathcal{Y}_L(h) = {\rm diag}\left(0,\, \mathcal{Y}_{E}(h)\right)\,,
    \\
    \mathcal{Y}_{U,D,E}(h) &= Y_{u,d,e}\left(1+ a_{u,d,e} \frac{h}{v}+ b_{u,d,e}\frac{h^2}{v^2}+...\right)\,.
\end{align}
The leading terms have been determined to obtain canonically normalized kinetic terms and to correctly match to the Higgs and fermions masses. In general, the terms $a_{u,d,e}$, $b_{u,d,e}$, ... are $3\times 3$ complex matrices in flavor space. For our purpose we will though only consider the coefficients associated to the top quarks, $a_t$ and $b_t$.
This parameterization allows us to obtain the SM by explicitly setting $a_{\lambda_3}=a_{u,d,e}=b_{\lambda_3}=b_C=1$, $a_C=2$, and $b_{u,d,e}=0$, which means that in the SM limit each parameter has to be tuned by hand.
\subsection{HEFT power counting}
As a final missing ingredient, we need to establish the power counting in order to systematically obtain higher orders in the EFT expansion. To this end, we will review the power counting proposed in Ref.~\cite{Brivio:2025yrr}, which will be applied in the following sections to Higgs pair production in gluon fusion. A consistent power counting is a cornerstone of any EFT expansion, as it prescribes where to truncate the series. 
\par
The guiding principle we adopt here is that of \textit{chiral dimension}~\cite{Buchalla:2013eza}, which provides a consistent framework to assign orders in the HEFT expansion up to the level of cross sections \cite{Brivio:2025yrr}.
Starting with naive dimensional analysis (NDA) one finds 
that a generic interaction on the Lagrangian level can be written as \cite{Manohar:1983md, Gavela:2016bzc}
\begin{equation}
 \Lag \supset \frac{\Lambda^4}{(4\pi)^2}
 \left(\frac{\de_\mu}{\Lambda}\right)^q   
 \left(\frac{4\pi\phi}{\Lambda}\right)^s   
 \left(\frac{4\pi\psi}{\Lambda^{3/2}}\right)^f   
 \left(\frac{g}{4\pi}\right)^{n_g}
 \left(\frac{\lambda}{(4\pi)^2}\right)^{n_\lambda}
 \left(\frac{4\pi v}{\Lambda}\right)^{n_v},
\label{eq:NDA}
\end{equation}
where each quantity in brackets is dimensionless. The fields $\phi$ stand for any scalar or gauge field insertion, whereas $\psi$ is a generic fermion field. A non-zero vacuum expectation value (VEV) $v$ has the same dimension  as a scalar field and the coupling constants $g$ and $\lambda$ are defined from the interactions $\phi \bar\psi\psi$ (where again $\phi$ stands for a scalar or vector) and the scalar self-interaction $\phi^4$ respectively.
While in SMEFT the NDA power counting allows to trivially sort by powers of the suppression scale $\Lambda$, which at the Lagrangian level corresponds to the canonical dimension, the power counting for HEFT is more complex. An important reason is that while in SMEFT one can expand in $(4\pi v/\Lambda)<1$, in HEFT this is no longer possible.  

We define the \textit{chiral dimension} to count the suppression factors in terms of $1/\Lambda$ or $1/4 \pi$, where $N_{\Lambda}$ and $N_{4 \pi}$ count the inverse powers, as~\cite{Brivio:2025yrr}
\begin{equation}
N_{\chi}=N_{\Lambda}+N_{4 \pi}\,.
\end{equation}
The chiral order of a Lagrangian interaction is given by
\[
N_{\chi,i}= -2 +q_i+\frac{f_i}{2}+N_{\chi_i}^{\kappa} \, , \label{eq:Nchii}
\]
where $q_i$ is the number of derivatives and $f_i$ is the number of fermion fields. $N_{\chi_i}^{\kappa}$ is the chiral dimension of the coupling, where  the strong, electroweak and Yukawa couplings contribute with $N_{\chi_i}^{\kappa}=1$ and the scalar self-coupling $\lambda$ with $N_{\chi_i}^{\kappa}=2$.

For details on how to arrive from a Lagrangian interaction to the cross section level, we refer to Ref.~\cite{Brivio:2025yrr}.
In practice, it is useful to have at hand a rule that allows to easily see whether a certain Feynman diagram contributes to the desired order of the HEFT expansion. As has been shown in Ref.~\cite{Brivio:2025yrr}, there are two possible consistent countings, either in 
\begin{align}
\label{N_HEFT}
N_{\rm HEFT}^{\mathcal{M}} =n-2+2 L +\sum_{i \in \text{vert}} (N_{\chi, i}-N_{g_s,i})
\end{align}
or in
\begin{align}
\label{Ns_HEFT}
N^{s,\mathcal{M}}_{\rm HEFT}=N_{\rm HEFT}^{\mathcal{M}} +N_{g_s}\,,
\end{align}
where $N_{g_s}$ is the number of strong coupling constants $g_s$ in the matrix element and $N_{g_s,i}$ in each vertex.
The reason why there are this two distinct possibilities for the counting is that while the electroweak and Yukawa couplings are related to the particle masses, which in turn need to be counted since they appear in propagators on the same footing than momenta, for the strong coupling constant, this is not the case and one is free to choose whether to include its powers into the counting or not.
Its inclusion somewhat streamlines the power counting as interaction terms from the same Lagrangian interaction (\eg\ all the various terms in the field strength $G_{\mu\nu}$) all count the same, while the expansion in $N_{\text{HEFT}}^{\mathcal{M}}$ decouples the perturbative expansion in the strong coupling constant from the EFT expansion, which allows to consider the two expansions separately. 
This possibility can be convenient in presence of large QCD corrections which can then be accounted for separately.
Finally, we note that both $N^{s,\mathcal{M}}_{\rm HEFT}$ and $N^{\mathcal{M}}_{\rm HEFT}$ depend on the chiral dimension of the interactions at each vertex as given in Eq.~\eqref{eq:Nchii}, the number of loops $L$ and the number of external legs $n$. The latter is a necessary ingredient to ensure an EFT expansion consistent with the cancellation of eventual IR divergencies, see Ref.~\cite{Brivio:2025yrr}. 
\par
In the remainder of this paper, we adopt the power counting defined by $N^{s,\mathcal{M}}_{\text{HEFT}}$. In this scheme, increasing the loop order of a process also requires the inclusion of diagrams with lower loop order but vertices of higher $N_{\chi,i}$. Consequently, higher-order perturbative corrections to the LO HEFT Lagrangian (such as QCD corrections) must be accompanied by operators that first appear at NLO or even NNLO in the HEFT Lagrangian.
For comparison, the approach in which NLO QCD corrections are added on top of the LO HEFT Lagrangian (see Refs.~\cite{Grober:2015cwa, Buchalla:2018yce}) corresponds instead to the $N^{\mathcal{M}}_{\text{HEFT}}$ counting, where the QCD corrections and the HEFT expansion are treated as two independent expansions.

\section{Double Higgs Production}
\label{sec_Di_Higgs_prod}
\begin{figure}[t] \centering
    \begin{minipage}{0.5\textwidth}
        \centering
    \raisebox{0cm}{\includegraphics[width=0.79\textwidth]{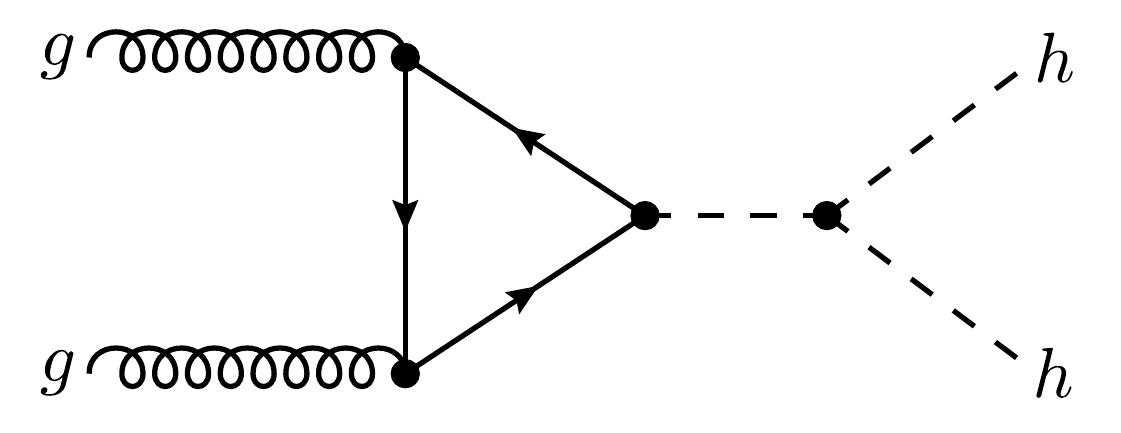}}
    \end{minipage}\hfill
    \begin{minipage}{0.5\textwidth}
        \centering
    \raisebox{0cm}{\includegraphics[width=0.77\textwidth]{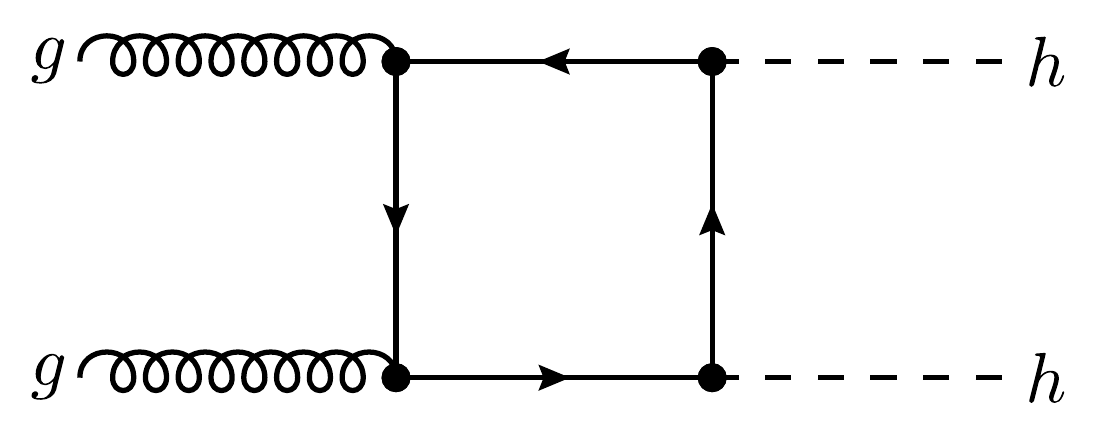}}
    \end{minipage}
\caption{Diagrammatic topologies contributing to $gg \rightarrow hh$ in the SM at LO.}
\label{diagrams_ggHH_SM}
\end{figure}

For our calculation of the Higgs pair production process $pp \rightarrow hh$ at proton-proton colliders in the HEFT and SMEFT frameworks, we restrict ourselves to the gluon-initiated process $gg \rightarrow hh$, which constitutes the dominant contribution to the cross section \cite{DiMicco:2019ngk}. Furthermore, we only retain the top quark and neglect contributions from bottom quark loops as their effect is below the $1\%$ level in the SM at LO \cite{Baglio:2012np}. In general, the amplitude for the partonic $gg \rightarrow hh$ subprocess can be written in terms of two projection operators forming a basis of the most general Lorentz structure the amplitude can attain \cite{Glover:1987nx}:
\begin{align}
\label{amplitude_form_factors}
    \mathcal{M}(g_a g_b \rightarrow hh) = \delta^{ab} \left(\mathcal{M}_1 \, A_1^{\mu \nu} + \mathcal{M}_2 \, A_2^{\mu \nu}\right) \, \epsilon_{\mu} \epsilon_{\nu}.
\end{align}
Defining $p_1, p_2$ ($p_3, p_4$) as the momenta of the incoming gluons (outgoing Higgs bosons), the explicit formulae for the projection operators expressed in terms of 
the square of the transverse momentum 
\begin{align}
    p_T^2 = \frac{2 (p_1 \cdot p_3)(p_2 \cdot p_3)}{(p_1 \cdot p_2)} - p_3^2
\end{align}
are given by
\begin{equation}
\begin{aligned}
    A_1^{\mu \nu} &= \frac{1}{\sqrt{2}} \left( g^{\mu \nu} - \frac{p_1^{\nu} p_2^{\mu}}{p_1 \cdot p_2}\right), \\
    A_2^{\mu \nu} &= \frac{1}{\sqrt{2}} \left( 
    g^{\mu \nu} + \frac{p_3^2 p_1^{\nu} p_2^{\mu}}{p_T^2 (p_1 \cdot p_2)} - 
    \frac{2 (p_2 \cdot p_3) p_1^{\nu} p_3^{\mu}}{p_T^2(p_1 \cdot p_2)} - \frac{2 (p_1 \cdot p_3) p_3^{\nu} p_2^{\mu}}{p_T^2(p_1 \cdot p_2)} + \frac{2 p_3^{\mu} p_3^{\nu}}{p_T^2}
    \right).
\end{aligned}
\end{equation}
The normalization of the projectors is chosen such that they form an orthonormal basis. The total partonic cross section $\hat{\sigma}$ of the process can then be cast into the form
\begin{align}
    \hat{\sigma}(gg \rightarrow hh) = \int_{\hat{t}_{-}}^{\hat{t}_+} d\hat{t} \ \frac{1}{1024 \pi \hat{s}^2} \left[|\mathcal{M}_1|^2 + |\mathcal{M}_2|^2 \right]
\end{align}
with $\hat{t}$ the Mandelstam variable characterizing the momentum transfer and integration boundaries given by
\begin{align}
    \hat{t}_{\pm} = - \frac{\hat{s}}{2} \left(1 \mp \sqrt{1- \frac{4 m_h^2}{\hat{s}}} \right) + m_h^2 \, .
\end{align}
The hadronic cross section at a center-of-mass (COM) energy $\sqrt{s}$ can then be obtained by convoluting $\hat{\sigma}$ with parton luminosity $\mathcal{L}_{gg}$ as
\begin{align}
    \sigma (pp \rightarrow hh) = \int_{4 m_h^2/s}^{1} d \tau \ \frac{d \mathcal{L}_{gg}}{d \tau} \ \hat{\sigma}(gg \rightarrow hh) \, ,
\end{align}
where we defined
\begin{align}
    \hat{s}=x_1 x_2 s & \,, & \tau & = x_1 x_2 \,, & \frac{d \mathcal{L}_{gg}}{d \tau} & = \int_{\tau}^{1} \frac{dx}{x} \, f_g(x, \mu_F) \, f_g(\tau/x, \mu_F)
\end{align}
with $f_g$ denoting the gluon parton distribution function (PDF) evaluated at a factorization scale $\mu_F$. In case of the LO SM calculation, two one-loop topologies contribute to double Higgs production, see Fig. \ref{diagrams_ggHH_SM}. In what follows, we refer to the triangle (box) graph form factors as $F^{\mathrm{SM}}_i$ ($G^{\mathrm{SM}}_i$). Using this notation, one finds
\begin{align}
    \mathcal{M}_1 = F^{\mathrm{SM}}_1 + G^{\mathrm{SM}}_1, \quad \quad \quad \mathcal{M}_2 = G^{\mathrm{SM}}_2,
\end{align}
where the explicit analytical expressions of $F^{\mathrm{SM}}_1, G^{\mathrm{SM}}_1$, and $G^{\mathrm{SM}}_2$ can be found in \cite{Glover:1987nx, Plehn:1996wb}. At NLO QCD no analytic formulae in full mass dependence are available, but they have been computed numerically \cite{Borowka:2016ypz, Borowka:2016ehy, Baglio:2020ini} or in analytic approximations, as for instance by using a small $p_T$ expansion combined with a high energy expansion \cite{Bonciani:2018omm, Davies:2018qvx, Bellafronte:2022jmo, Davies:2023vmj}. We will make use of the NLO QCD corrections as computed in \cite{Buchalla:2018yce} for the LO HEFT Lagrangian, based on the numeric calculation of the NLO QCD corrections in \cite{Borowka:2016ehy}. Finally, we note that a division in box and triangle topologies strictly speaking is no longer possible at NLO in EFT or pertubation theory, but we only need to characterize contributions in terms of couplings and effective operators.
In the following section, we discuss how to extend the SM calculation to the HEFT framework by consistently including with the QCD corrections also higher orders in the EFT expansion.

\subsection{HEFT Calculation} \label{Sec_HEFT_Calculation}
For our HEFT di-Higgs analysis, we adopt the power counting scheme based on the chiral dimension $N_{\chi}$ as introduced in Section \ref{sec_powercounting_recap}. The truncation of cross section contributions that arise from the product of matrix elements $\mathcal{M}_a$ and $\mathcal{M}_b$ is performed by selecting a maximal order $N_{\rm HEFT}^{\mathrm{trunc}}$ and only keeping contributions that satisfy
\begin{align}
    N_{\rm HEFT}^{s} = N_{\rm HEFT}^{s, a} + N_{\rm HEFT}^{s, b} \leq N_{\rm HEFT}^{\mathrm{trunc}}.
\end{align}
In this work, we include contributions up to $N_{\rm HEFT}^{\mathrm{trunc}} = 10$, which is motivated by the state-of-the-art calculation of Ref.~\cite{Buchalla:2018yce}, which captures the effect of NLO QCD corrections of the contributions involving $N_{\chi}=0$ HEFT operators and the $N_{\chi}=2$ operator $g_s^2 G^a_{\mu \nu} G^{a \mu \nu} \mathcal{F}_G(h)$. As will be displayed in Table~\ref{tab:classification}, the two-loop and real radiation pieces scale as $N_{\rm HEFT}^{s, \mathcal{M}} = 6$ and $N_{\rm HEFT}^{s, \mathcal{M}} = 5$, respectively, such that the LO-virtual interference and the square of the real radiation part give rise to cross section contributions with $N_{\rm HEFT}^{s} = 10$. However, confronting the di-Higgs calculation with the $N_{\chi}=2$ and $N_{\chi}=4$ operators that have been reported in Refs.~\cite{Buchalla:2013rka,Brivio:2016fzo,Sun:2022ssa, Sun:2022snw}, the NLO QCD corrections are not the only contributions that arise up to $N_{\rm HEFT}^{\mathrm{trunc}} = 10$. Thus, we aim to complement the results of Ref.~\cite{Buchalla:2018yce} by the missing contributions that stem from these $N_{\chi}=2$ and $N_{\chi}=4$ operators.
In particular, we focus on the CP and custodial symmetry preserving operators and assume a weakly coupled and renormalizable UV completion. Thus, each insertion of a gluon field strength tensor $G^{a}_{\mu \nu}$ features one power of $g_s$ and chirality flips involving top quark fields are weighted by a power of the top Yukawa coupling $y_t$, assuming minimal flavour violation (MFV) \cite{DAmbrosio:2002vsn}. This UV assumption effectively increases the chiral dimension of the effective operators and leads to a shift towards a higher perturbative order in the weakly coupled scenario. The corresponding set of operators, together with their $N_{\Lambda}, N_{4 \pi}$, and $N_{\chi}$ assignments are shown in Table \ref{operators_di_higgs_heft}.
\begin{table}[t]\centering
\renewcommand{\arraystretch}{1.3}
\begin{tabular}{c|l|c|c|c}
\hline
\textbf{operator} & \textbf{parameters}& $N_{\Lambda}$ & $N_{4\pi}$& $N_{\chi}$  \\\hline
$G_{\mu \nu}^a G^{a \mu \nu}$ & & $0$ & 0 & 0\\\hline
$\lambda v^4\mathcal{V}(h)$ & $a_{\lambda^3}$ & $0$ & 0 & $0$ \\\hline
$i \overline{Q}_L \slashed{D} Q_L$  & & 0 & 0 & $0$ \\\hline
$\overline{Q}_L \mathbf{U} \mathcal{Y}_Q(h) Q_R$ & $a_t, b_t$& 0 & 0 & $0$ \\\hline
$g_s^2 G_{\mu \nu}^a G^{a \mu \nu} \mathcal{F}_G(h)$ & $a_g, b_g$ & 0 & 2 & $2$ \\\hline
$g_s y_t \overline{Q}_L \sigma^{\mu \nu} G_{\mu \nu}^a T^a \mathbf{U} Q_R \mathcal{F}_{C}(h)$ &$d_c, a_c, b_c$ &1 & 1 & $2$ \\\hline
$y_t (\partial_{\mu} h  \partial^{\mu} h) \overline{Q}_L \mathbf{U} Q_R \mathcal{F}_{D}(h)$& $b_D$ & 3 & $-1$ & $2$ \\\hline
$g_s^2 h^2 (D^{\mu} G^{a \nu \lambda}) (D_{\mu} G_{\nu \lambda}^a) \mathcal{F}_{G1}(h)$ & $b_g^{(1)}$ & 4 & 0 & $4$ \\\hline
$g_s^2 h G^{a \lambda \nu} G_{\lambda}^{a \mu} (\partial_{\mu} \partial_{\nu} h) \mathcal{F}_{G2}(h)$ & $b_g^{(2)}$ & 4 & 0 & $4$ \\\hline
\end{tabular}
\caption{Classification of HEFT operators up to $N_{\chi} = 4$ contributing to di-Higgs production. The second column reports the relevant parameters for $gg\to hh$.}
\label{operators_di_higgs_heft}
\end{table}
Expanding out all of the $\mathcal{F}(h)$ functionals in powers in the physical Higgs field $h$ and extracting the explicit factors of $\Lambda$ and $4 \pi$ to dimensionally normalize the Wilson coefficients, we obtain the following terms in the HEFT Lagrangian:
\begin{equation}
\begin{aligned}
\label{Lagrangian_Di_Higgs_HEFT}
    \mathcal{L}_{\mathrm{HEFT}} &\supset \mathcal{L}_{\kappa} + \delta\mathcal{L} \\
    \mathcal{L}_{\kappa} &=  \frac{1}{2} (\partial_{\mu} h)^2 - \frac{1}{2} m_h^2 h^2 - a_{\lambda^3} \lambda v h^3
    - \frac{1}{4} G_{\mu \nu}^a G^{a \mu \nu} + i \bar{t} \slashed{D} t - \frac{y_t v}{\sqrt{2}} \left(a_t \frac{h}{v} + b_t \frac{h^2}{v^2}\right) \overline{t} t \\
    &+ \frac{g_s^2}{16 \pi^2} G_{\mu \nu}^a G^{a \mu \nu} \left(a_g \frac{h}{v} + b_g \frac{h^2}{v^2} \right) \\
    \delta\mathcal{L} &= \frac{y_t b_D}{4 \pi \Lambda} \frac{1}{v^2} (\partial_{\mu} h)^2 \overline{t} t
    + \frac{g_s y_t}{4 \pi \Lambda} \left(\overline{t}_L \sigma^{\mu \nu} G_{\mu \nu}^a T^a t_R + \mathrm{h.c.} \right) \left(d_c + a_c \frac{h}{v} + b_c \frac{h^2}{v^2}\right) \\
    &+ \frac{g_s^2 b_{g}^{(1)}}{16 \pi^2 \Lambda^2} \frac{h^2}{v^2} (D^{\mu} G^{a \nu \lambda}) (D_{\mu} G_{\nu \lambda}^a) + \frac{g_s^2 b_{g}^{(2)}}{16 \pi^2 \Lambda^2} \frac{h}{v} G^{a \lambda \nu} G_{\lambda}^{a \mu} \frac{1}{v} (\partial_{\mu} \partial_{\nu} h) \, . 
\end{aligned}
\end{equation}
Here, $\mathcal{L}_{\kappa}$ corresponds to the well-studied $\kappa-$formalism Lagrangian that introduces multiplicative modifications of the SM couplings, together with an $h^2 \bar{t} t$ term and Higgs-gluon contact interactions. We note that most of the operators are of order $N_{\chi}=0$, except for the operator that couples the gluon field strength to the Higgs bosons, which is of order $N_{\chi}=2$. For its phenomenological relevance, it usually is included in $\mathcal{L}_{\kappa}$. Since we intend to compare with previous works, we define $\mathcal{L}_{\kappa}$ accordingly. The additional operators in $\delta \mathcal{L}$ represent the missing $N_{\chi} = 2$ and $N_{\chi} = 4$ contributions required when working up to $N_{\mathrm{HEFT}}^{s,\mathcal{M}}=6$. Note, that the only feasible direct probe of the coefficients involving two Higgs bosons, namely $\{b_t, b_D, b_c, b_g^{(1)}, b_g^{(2)} \}$, is the production of two Higgs bosons. Having now the relevant terms of the Lagrangian $\mathcal{L}_{\mathrm{HEFT}}$ at hand, we can revisit Eqs.~(\ref{N_HEFT}) and (\ref{Ns_HEFT}) to determine which operators can enter the Feynman diagrams up to which loop order $L$ to consistently count in the parameter $N^{s,\mathcal{M}}_{\text{HEFT}}$. The classification of the Feynman diagrams according to the loop order $L$ and the $N_{\chi}$ of the operators can be found in Table \ref{tab:classification}.
\begin{table}
\begin{center}
\begin{tabular}{c|c|ccc|c}
$n$&
$L$&
\parbox{1.5cm}{\centering $N_\chi=0$\\[-1mm] insertions}& 
\parbox{1.5cm}{\centering $N_\chi=2$\\[-1mm] insertions}& 
\parbox{1.5cm}{\centering $N_\chi=4$\\[-1mm] insertions}& $N_{\rm HEFT}^{s, \meLabel}$
\\[3mm]\hline
\rowcolor{SkyBlue!20}
4&0& any& 1& -& 4
\\
\rowcolor{SkyBlue!20}
4&1& any& -& -& 4
\\
\rowcolor{SkyBlue!20}
5&0& any& 1& -& 5
\\
\rowcolor{SkyBlue!20}
5&1& any& -& -& 5
\\
4&0& any& -& 1& 6
\\
\rowcolor{SkyBlue!20}
4&0& any& 2& -& 6
\\
4&1& any& 1& -& 6
\\
\rowcolor{SkyBlue!20}
4&2& any& -& -& 6
\\\hline
\end{tabular} 
\caption{Classification of contributions to Higgs pair production up to $N_{\rm HEFT}^{s, \meLabel}=6$. The subset of diagrams highlighted in blue corresponds to results already obtained in Ref.~\cite{Buchalla:2018yce} that considers the $\mathcal{L}_{\kappa}$ Lagrangian. Real radiation diagrams have $n=5$. \label{tab:classification}}
\end{center}
\end{table}
In our selection of operators and diagrams, we consistently omit all one-loop electroweak corrections that would appear within our power counting scheme. Examples include loop diagrams involving the $G^a_{\mu \nu} G^{a\mu \nu} h^3$ operator and electroweak corrections to the trilinear Higgs self-coupling as considered in \cite{Englert:2025xrc}. A full treatment of all the electroweak effects requires the computation of two-loop multi-scale higher order corrections, and is beyond the scope of this work. We refer to \cite{Bi:2023bnq} for the full electroweak corrections to double Higgs production in the SM and to \cite{Bizon:2018syu, Borowka:2018pxx} for partial corrections in the $\kappa_{\lambda}$ framework. On top of that, we also exclude four-fermion operators, which can attain the schematic forms of $t_L^2 t_R^2, t_L^4,t_R^4$, and enter at two-loop order. Tree-level matching results for the SMEFT show that $t_L^2 t_R^2$ operators typically arise with powers of $y_t^2$ \cite{deBlas:2017xtg}.
The respective insertion of four-top operators into two-loop topologies have been computed in Ref.~\cite{Heinrich:2023rsd}, corresponding to diagrams of order $N_{\rm HEFT}^{s,\meLabel}=8$ if $y_t^2$ is factored out and hence beyond the order we consider. For four-fermion operators involving only the same chirality instead, we note that as shown in Ref.~\cite{Heinrich:2023rsd}, same chirality four-top operators have a very suppressed contribution to di-Higgs production\footnote{Operators of type $t_L^2 t_R^2$ have significantly larger contributions but are subject to the scheme choice for $\gamma_5$ \cite{DiNoi:2023ygk, Heinrich:2023rsd}.}, motivating why we consider their contribution to be beyond the order we consider in this work.
\par
\begin{figure}[t] \centering
    \begin{minipage}{0.33\textwidth}
        \centering
        \includegraphics[width=1\textwidth]{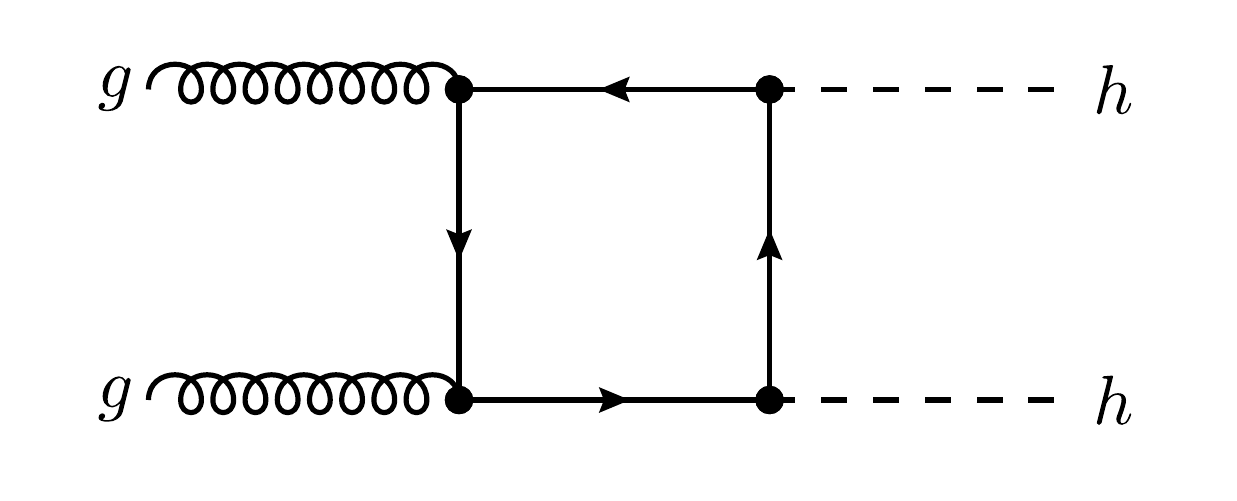}
        \subcaption{$L=1$, $N_{\chi}=0$ insertions}
    \end{minipage}
        \begin{minipage}{0.33\textwidth}
        \centering
        \includegraphics[width=0.6\textwidth]{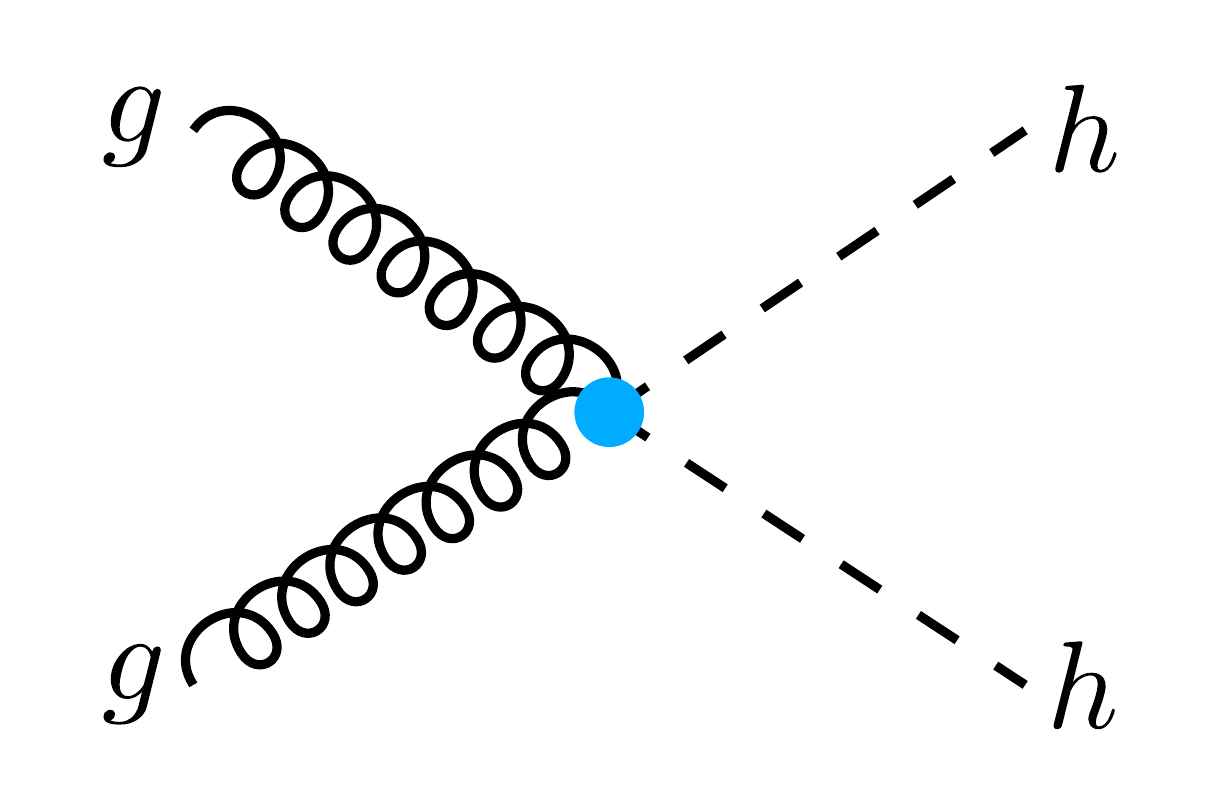}
        \subcaption{$L=0$, $N_{\chi}=2$ insertion}
    \end{minipage}
    \begin{minipage}{0.33\textwidth}
        \centering
    \includegraphics[width=1\textwidth]{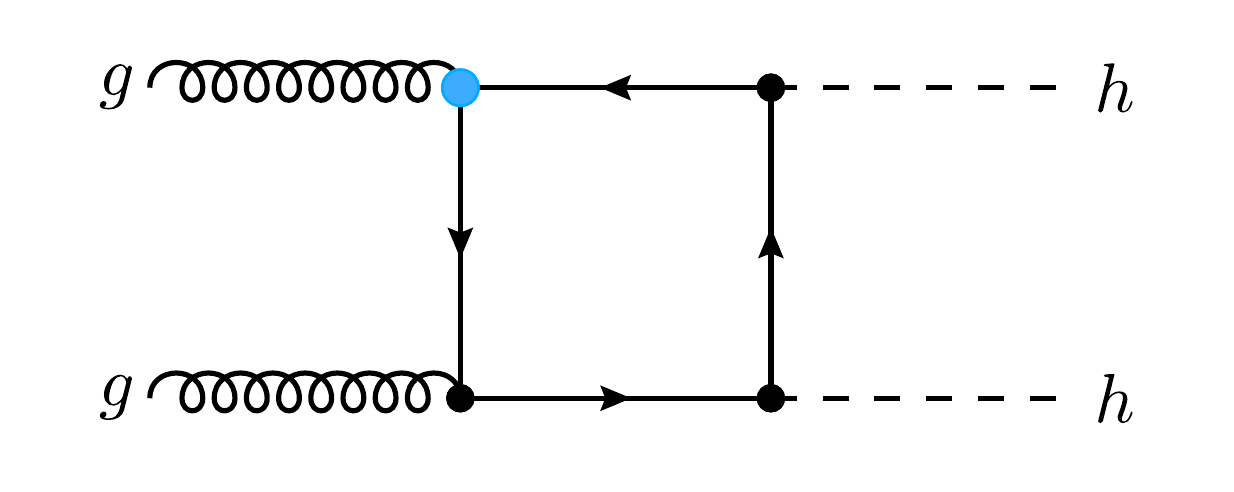}
    \subcaption{$L=1$, $N_{\chi}=2$ insertion}
    \end{minipage}\hfill
    \begin{minipage}{0.33\textwidth}
        \centering
        \includegraphics[width=0.6\textwidth]{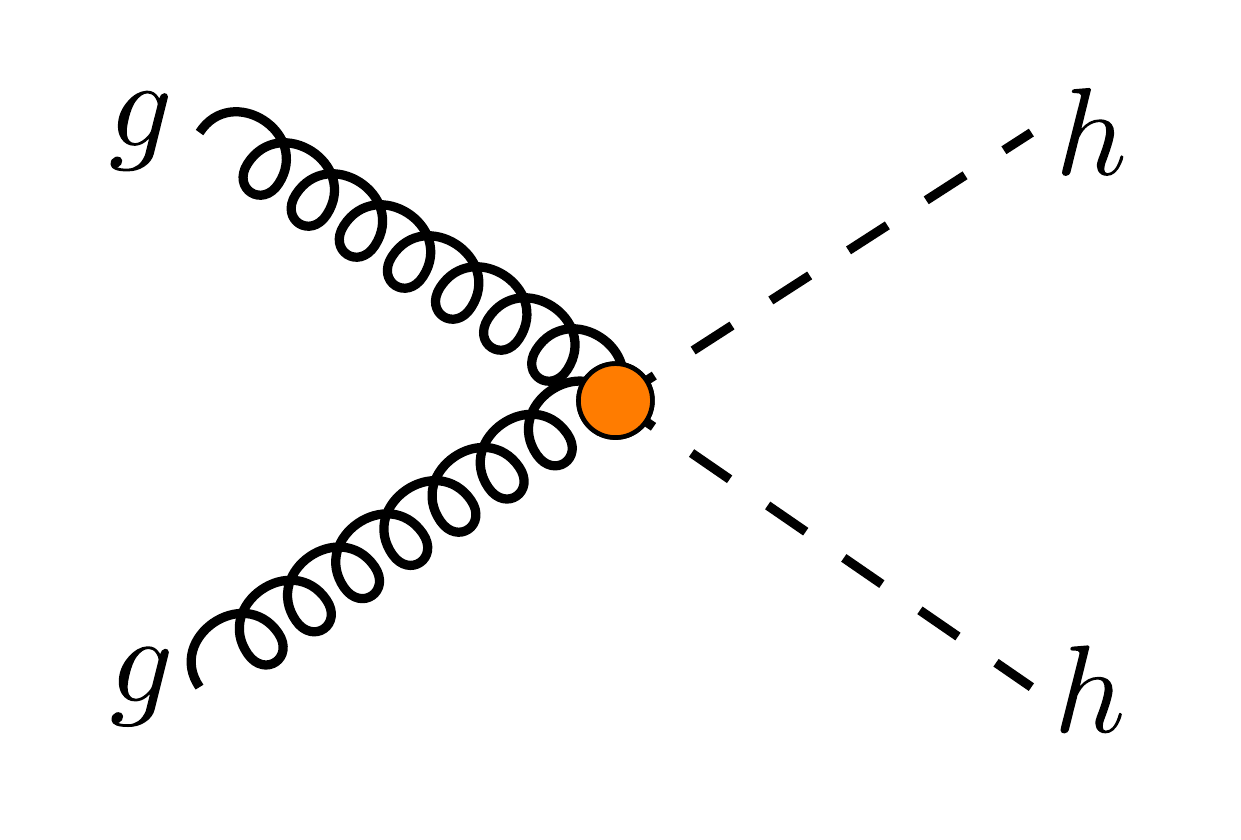}
        \subcaption{$L=0$, $N_{\chi}=4$ insertion}
    \end{minipage}
        \begin{minipage}{0.33\textwidth}
        \centering
        \includegraphics[width=0.75\textwidth]{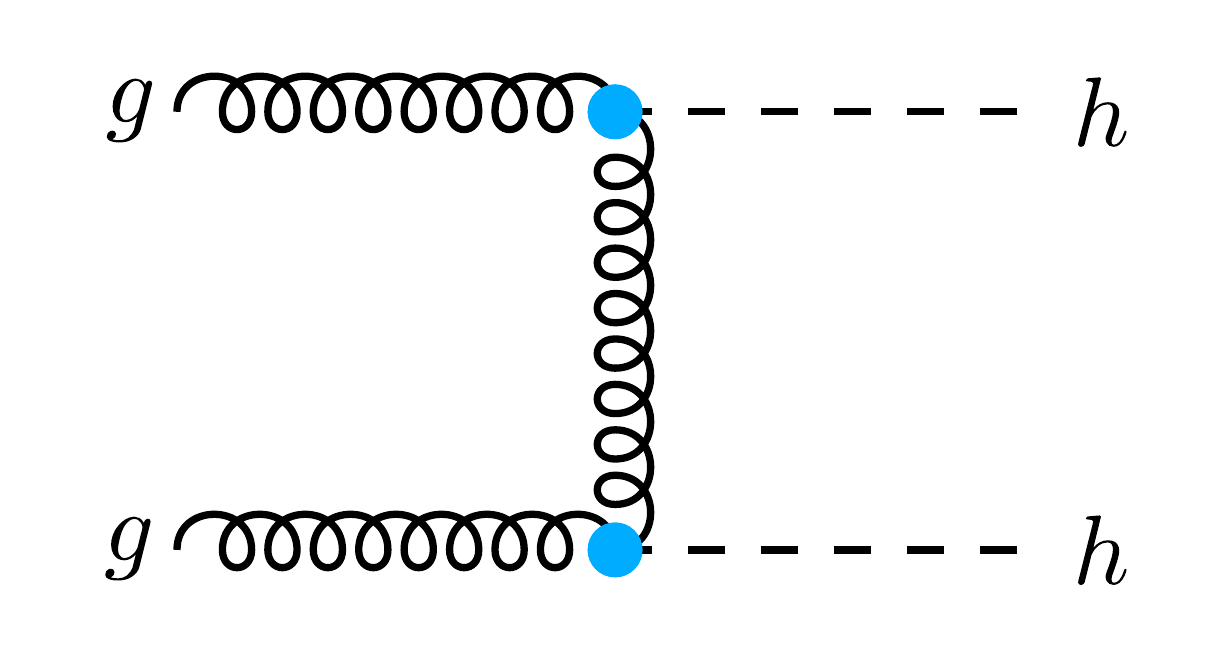}
        \subcaption{$L=0$, two $N_{\chi}=2$ insertions}
    \end{minipage}
\caption{Representative diagrams contributing to di-Higgs production up to $N_{\rm HEFT}^{s, \meLabel} = 6$. Blue (orange) dots indicate the insertion of a $N_{\chi} = 2$ ($N_{\chi} = 4$) operator.}
\label{Di_Higgs_HEFT_diagrams}
\end{figure}
\begin{figure}[t] \centering
\includegraphics[width=0.8\textwidth]{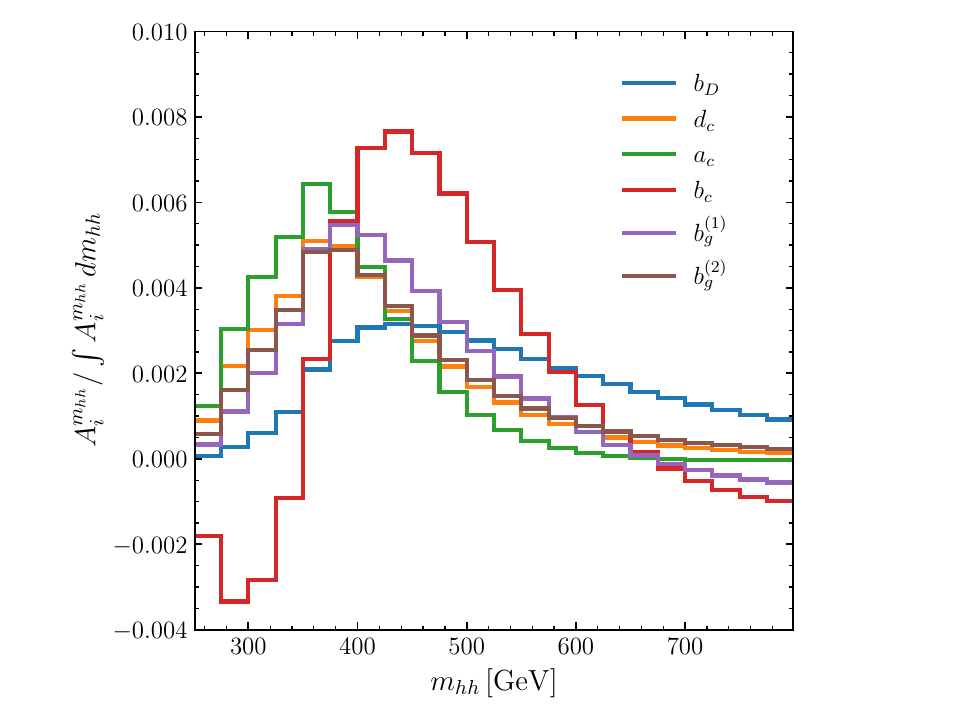}
\caption{Normalized bin-wise coefficients $A^{m_{hh}}_i$ that change the SM invariant mass distributions in presence of the EFT coefficients $\{b_D, d_c, a_c, b_c, b_g^{(1)}, b_g^{(2)}\}$.}
\label{figure_new_coefficients_SM_kappas}
\end{figure}

Some representative Feynman diagrams of contributions that we include in our calculation, together with their $N_{\chi}$ scaling are displayed in Fig. \ref{Di_Higgs_HEFT_diagrams}. The contributions highlighted in blue in Table~\ref{tab:classification} have been evaluated in Ref.~\cite{Buchalla:2018yce}, where the authors studied the contribution to di-Higgs production of the $\mathcal{L}_{\kappa}$ Lagrangian up to NLO QCD in the chiral counting and we extend these results by calculating the missing contributions originating from $\delta \mathcal{L}$. For convenience, we parameterize the di-Higgs production cross section, differential in the invariant mass $m_{hh}$ of the Higgs pair, as 
\begin{align}
\label{parameterization_invariant_mass}
    \frac{d \sigma}{d m_{hh}} = \sum_{i} A_i(m_{hh}) \, c_i \,,
\end{align}
where the $A_i$ denote bin-wise constants of unit $\mathrm{fb}/\mathrm{GeV}$ that depend on the central value of the invariant mass bin, whereas the $c_i$ are different combinations of the HEFT coefficients. Truncating the cross section contributions at HEFT order $N_{\rm HEFT}^{\mathrm{trunc}} = 10$ yields $64$ different $A_i$ factors, of which $A_1 - A_{23}$ were computed in \cite{Buchalla:2018yce}. 
\newpage
\noindent The full expression reads:
\begin{equation}
\begin{aligned}
\label{dsigma_dmhh_HEFT}
    \frac{d \sigma}{d m_{hh}} &= A_1 a_t^4 + A_2 b_t^2 + A_3 a_t^2 a_{\lambda^3}^2 + A_4 a_g^2 a_{\lambda^3}^2 + A_5 b_g^2 + A_6 b_t a_t^2 + A_7 a_t^3 a_{\lambda^3} \\
    &+ A_8 a_t b_t a_{\lambda^3} + A_9 b_t a_g a_{\lambda^3} + A_{10} b_t b_g + A_{11} a_t^2 a_g a_{\lambda^3} + A_{12} a_t^2 b_g \\
    &+ A_{13} a_t a_{\lambda^3}^2 a_g + A_{14} a_t a_{\lambda^3} b_g + A_{15} a_g a_{\lambda^3} b_g + A_{16} a_t^3 a_g + A_{17} a_t b_t a_g \\
    &+ A_{18} a_t a_g^2 a_{\lambda^3} + A_{19} a_t a_g b_g + A_{20} a_t^2 a_g^2 + A_{21} b_t a_g^2 + A_{22} a_g^3 a_{\lambda^3} + A_{23} a_g^2 b_g \\
    &+ A_{24} b_D a_t a_{\lambda^3} + A_{25} b_D b_t + A_{26} b_D a_t^2 + A_{27} b_D a_g a_{\lambda^3} + A_{28} b_D b_g \\
    &+ A_{29} d_c a_t^2 a_{\lambda^3}^2 + A_{30} d_c a_t^3 a_{\lambda^3} + A_{31} d_c a_t b_t a_{\lambda^3} + A_{32} d_c b_t a_t^2 + A_{33} d_c b_t^2 \\
    &+ A_{34} d_c a_t^4 + A_{35} d_c a_g a_t a_{\lambda^3}^2 + A_{36} d_c a_g a_t^2 a_{\lambda^3} + A_{37} d_c a_g b_{t} a_{\lambda^3} \\
    &+ A_{38} d_c b_g a_t a_{\lambda^3} + A_{39} d_c b_g a_t^2 + A_{40} d_c b_g b_t + A_{41} a_c a_t a_{\lambda^3}^2 \\
    &+ A_{42} a_c a_t^2 a_{\lambda^3} + A_{43} a_c b_t a_{\lambda^3} + A_{44} a_c a_t b_t + A_{45} a_c a_t^3 + A_{46} a_c a_G a_{\lambda^3}^2 \\
    &+ A_{47} a_c a_g a_t a_{\lambda^3} + A_{48} a_c b_g a_{\lambda^3} + A_{49} a_c b_g a_t + A_{50} b_c a_t a_{\lambda^3} + A_{51} b_c b_t  \\
    &+ A_{52} b_c a_t^2 + A_{53} b_c a_g a_{\lambda^3} + A_{54} b_c b_g + A_{55} b_{g}^{(1)} a_t a_{\lambda^3} + A_{56} b_{g}^{(1)} b_t \\
    &+ A_{57} b_{g}^{(1)} a_t^2 + A_{58} b_{g}^{(1)} a_g a_{\lambda^3} + A_{59} b_{g}^{(1)} b_g + A_{60} b_{g}^{(2)} a_t a_{\lambda^3} + A_{61} b_{g}^{(2)} b_t \\
    &+ A_{62} b_{g}^{(2)} a_t^2 + A_{63} b_{g}^{(2)} a_g a_{\lambda^3} + A_{64} b_{g}^{(2)} b_g \, ,
\end{aligned}
\end{equation}
\\ 
where $A_1-A_{15}$ comprise contributions $N_{\mathrm{HEFT}}^s=8, \,10$ and $A_{16}-A_{64}$ originate from $N_{\mathrm{HEFT}}^s=10$ contributions only. For our analysis of the novel $A_i$ coefficients, we use central bin values between $262.5\, \mathrm{GeV}$ and $787.5\, \mathrm{GeV}$ and a bin width of $25 \, \mathrm{GeV}$ as in \cite{Buchalla:2018yce}. The input values for the numerical analysis are chosen accordingly:
\begin{align}
    m_t = 173 \, \mathrm{GeV}, \ \ m_h = 125 \, \mathrm{GeV}, \ \ \alpha_s(m_Z) = 0.118, \ \ \sqrt{s} = 13\,\mathrm{TeV},
\end{align}
together with the parton distribution functions (PDFs) \texttt{PDF4LHC15\_nlo\_100\_pdfas} \cite{Butterworth:2015oua, Dulat:2015mca}. Divergences that arise from one-loop diagrams with HEFT operator insertions are treated using the $\overline{\mathrm{MS}}$ scheme. For the renormalization and factorization scales, we choose \mbox{$\mu_F = \mu_R \equiv \mu = m_{hh}/2$}, and the dimensionful scale $\Lambda$ is set to $1 \, \mathrm{TeV}$. The numerical values of $A_{24}-A_{64}$ in this setup can be found in an ancillary file.  \\ \\
In order to illustrate the effect of the EFT coefficients contained in $\delta \mathcal{L}$ on the SM distribution, we simplify the parameterization by taking the SM limit of $\{a_t, b_t, a_{\lambda^3}, a_g, b_g \}$, which allows us to write
\begin{align}
    \frac{d \sigma}{d m_{hh}} =  \frac{d\sigma_{\mathrm{SM}}^{\mathrm{NLO}}}{d m_{hh}} + A^{m_{hh}}_{b_D} \, b_D + A^{m_{hh}}_{d_c} \, d_c + A^{m_{hh}}_{a_c} \, a_c + A^{m_{hh}}_{b_c} \, b_c + A^{m_{hh}}_{b_g^{(1)}} \, b_g^{(1)} + A^{m_{hh}}_{b_g^{(2)}} \, b_g^{(2)},
\end{align}
where based on Eq.~(\ref{dsigma_dmhh_HEFT}) we identified
\begin{equation}
\begin{aligned}
\frac{d\sigma_{\mathrm{SM}}^{\mathrm{NLO}}}{d m_{hh}} 
&= A_1 + A_3 + A_7\, , 
& A_{b_D}^{m_{hh}} 
&= A_{24} + A_{26}\, , 
& A_{d_c}^{m_{hh}} 
&= A_{29} + A_{30} + A_{34}\, , \\
A_{a_c}^{m_{hh}} 
&= A_{41} + A_{42} + A_{45}\, , 
& A_{b_c}^{m_{hh}} 
&= A_{50} + A_{52}\,, 
& A_{b_g^{(1)}}^{m_{hh}} 
&= A_{55} + A_{57}\,, \\
A_{b_g^{(2)}}^{m_{hh}} 
&= A_{60} + A_{62}\,.
\end{aligned}
\end{equation}
The normalized $A^{m_{hh}}_i$ coefficients are shown in Fig.~\ref{figure_new_coefficients_SM_kappas}. Before turning to a detailed study of the invariant mass distributions in Section \ref{sec_invariant_mass}, we next derive positivity bounds on the HEFT Wilson coefficients.

\subsection{Positivity Bounds}
\label{sec_positivity}
\begin{figure}[t] \centering
\includegraphics[width=0.8\textwidth]{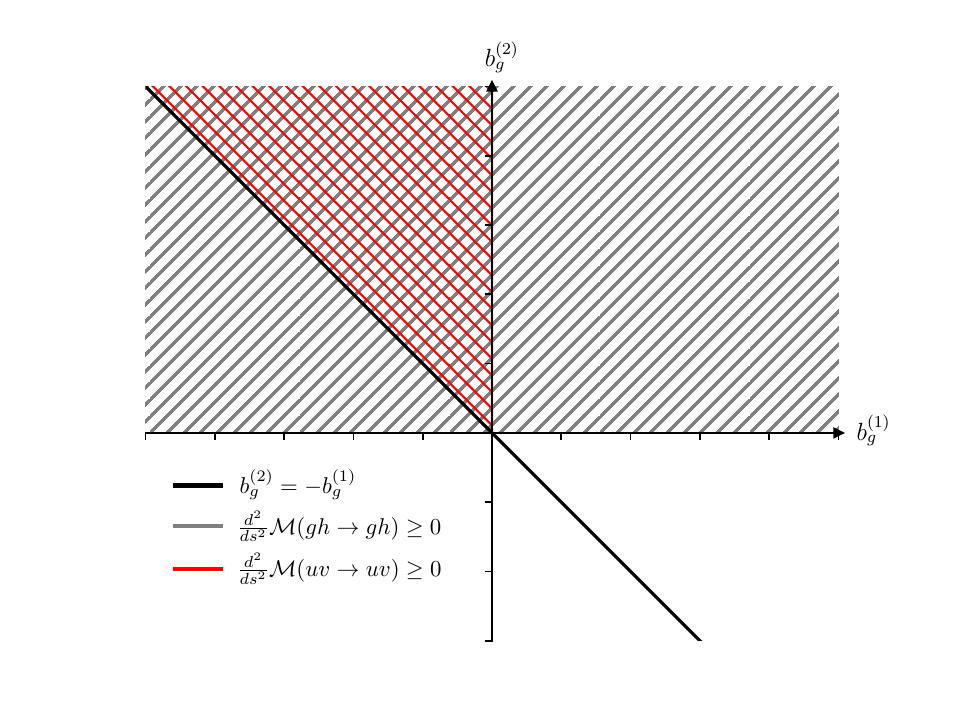}
\caption{Comparison of positivity bounds obtained using only the $gh \rightarrow gh$ process (gray shaded region) and the elastic $uv \rightarrow uv$ amplitude of the composite states $|u\rangle$ and $|v \rangle$ (red shaded region).}
\label{figure_positivity}
\end{figure}

To ensure compatibility of the Lagrangian in Eq.~(\ref{Lagrangian_Di_Higgs_HEFT}) with causality and the unitarity of the $S$-matrix \cite{Adams:2006sv, Zhang:2021eeo}, only a subset of signs and magnitudes of the a priori general HEFT Wilson coefficients is allowed. We derive these constraints following the approach of \cite{Bellazzini:2015cra, Cheung:2016yqr, deRham:2018qqo, Zhang:2021eeo, Chala:2023xjy}, by considering the two auxiliary superposition states $|u\rangle$ and $|v\rangle$ as
\begin{align}
    |u\rangle = u_h |h\rangle + \sum_{c, \lambda} u_{c \lambda} |g_{c \lambda}\rangle, \ \ \
    |v\rangle = v_h |h\rangle + \sum_{c, \lambda} v_{c \lambda} |g_{c \lambda}\rangle,
\end{align}
where $c$ and $\lambda$ denote the different colors and polarizations that the gluons can attain. Without loss of generality, we can take the coefficients $u_i$ and $v_i$ to be real. Furthermore, we assume that the states are normalized such that
\begin{align}
    u_h^2 + \sum_{c, \lambda} u_{c \lambda}^2 = 1, \ \ \ v_h^2 + \sum_{c, \lambda} v_{c \lambda}^2 = 1.
\end{align}
The optical theorem then implies that the second derivative with respect to the Mandelstam variable $s$ of the elastic scattering $uv \rightarrow uv$ amplitude is positive in the forward limit:
\begin{align}
\label{positivity_composite_amplitude}
    \frac{d^2}{ds^2} \mathcal{M} (uv \rightarrow uv) \geq 0.
\end{align}
When expanding out the composite states in this relation, only three of the partial amplitudes give a non-vanishing contribution:
\begin{equation}
\begin{aligned}
    \frac{d^2}{ds^2} \mathcal{M}(hh \rightarrow g_{c \lambda} \, g_{c' \lambda'}) &= \frac{g_s^2 b_g^{(1)}}{4 \pi^2 v^2 \Lambda^2} \delta^{c c'} \delta^{\lambda \lambda'}, \\
    \frac{d^2}{ds^2} \mathcal{M}(g_{c \lambda}\, g_{c' \lambda'} \rightarrow hh) &= \frac{g_s^2 b_g^{(1)}}{4 \pi^2 v^2 \Lambda^2} \delta^{c c'} \delta^{\lambda \lambda'},\\
    \frac{d^2}{ds^2} \mathcal{M}(g_{c \lambda} h \rightarrow g_{c' \lambda'} h) &= \frac{g_s^2 b_g^{(2)}}{8 \pi^2 v^2 \Lambda^2} \delta^{c c'} \delta^{\lambda \lambda'}.
\end{aligned}
\end{equation}
Hence, the positivity relation (\ref{positivity_composite_amplitude}) reduces to
\begin{align}
    \sum_{c, \lambda} \left[ u_{c \lambda} v_{c \lambda} u_h v_h \, \frac{g_s^2 b_g^{(1)}}{2 \pi^2 v^2 \Lambda^2} + (v_h u_{c \lambda} + u_h v_{c \lambda})^2  \frac{g_s^2 b_g^{(2)}}{8 \pi^2 v^2 \Lambda^2} \right]\geq 0,
\end{align}
which must hold for all possible values of the parameters $u_i$ and $v_i$. Introducing the two real parameters
\begin{align}
    R = \sum_{c, \lambda} u_{c \lambda} v_{c \lambda} u_h v_h, \ \ \ \tilde{R} = \sum_{c, \lambda} (v_h u_{c \lambda} + u_h v_{c \lambda})^2,
\end{align}
that satisfy $\tilde{R} \geq 0$ and $\tilde{R} > 4R$ for $R > 0$, the following cases lead to non-trivial positivity constraints for the Wilson coefficients $b_g^{(1)}$ and $b_g^{(2)}$:
\begin{equation}
\begin{aligned}
\label{relations_positivity}
    R = 0, \, \tilde{R} > 0&: \ \ b_g^{(2)} \geq 0, \\
    R < 0, \, \tilde{R} = 0&: \ \ b_g^{(1)} \leq 0, \\
    R > 0, \, \tilde{R} > 4R&: \ \ b_g^{(2)} \geq -b_g^{(1)}.
\end{aligned}
\end{equation}
Therefore, positivity fixes the sign of the coefficients $b_g^{(1)}$ and $b_g^{(2)}$, and further slices the $b_g^{(1)}-b_g^{(2)}-$plane by a linear function, see Fig. \ref{figure_positivity}. In our following numerical analysis of the invariant mass and angular distributions for the HEFT, we restrict the parameter space to those regions that satisfy the requirements listed in (\ref{relations_positivity}).

\subsection{Analysis of invariant mass distributions}
\label{sec_invariant_mass}
The inclusion of EFT operators on top of the SM prediction for a given process not only changes the total cross section, but may lead to new kinematic features on the level of differential distributions. These features may manifest as novel peak structures, shoulders, or modifications of the tail. In the following, we provide a comprehensive analysis of these shapes for the invariant mass distributions of the di-Higgs production process within both HEFT and SMEFT by employing a clustering algorithm that we describe below.

\subsubsection{Clustering Algorithm}
In recent literature, several clustering algorithms for the invariant mass distributions of di-Higgs production have been proposed \cite{Carvalho:2015ttv, Capozi:2019xsi}. Applied to the non-linear $\mathcal{L}_{\kappa}$ Lagrangian, a number of $N_{\mathrm{Clus}} = 8$ or $N_{\mathrm{Clus}} = 12$ representative clusters (including the SM distribution) has been identified to cover several different kinematic features that arise when varying the corresponding five coupling modifiers. As discussed in Section \ref{Sec_HEFT_Calculation}, a complete HEFT description also requires the consideration of the operators contained in  $\delta \mathcal{L}$, leading to six additional coefficients that can affect the differential shapes. In order to extend the previous analyses by the remaining degrees of freedom, we develop a clustering algorithm that compares binned distributions based on a $\chi^2-$test. In this article, we use the following notation for a point in the 11-dimensional parameter space of the HEFT coefficients
\begin{align}
    P = (a_{\lambda^3}, a_t, b_t, a_g, b_g, b_D, d_c, a_c, b_c, b_g^{(1)}, b_g^{(2)})
\end{align}
and label the corresponding normalized invariant mass distribution, derived from Eq. (\ref{dsigma_dmhh_HEFT}), as $D$. Given two sets of coefficients $P_1$ and $P_2$, we compare the shapes of $D_1$ and $D_2$ by computing their $\chi^2-$value: we assume a bin-dependent variance that can be computed from the relative uncertainty $\Delta_i$ that we apply equally to both of the normalized differential distributions as
\begin{align}
    \sigma_i^2=(\Delta_{1,i} D_{1,i})^2+(\Delta_{2,i} D_{2,i})^2=\Delta_i^2 (D_{1,i}^2+D_{2,i}^2),
\end{align}
leading to
\begin{align}
\label{chi2_general_formula}
    \chi^2(P_1, P_2) = \sum_{i \in \mathrm{bins}} \frac{(D_{1, i} - D_{2, i})^2}{\Delta_i^2 (D_{1, i}^2 + D_{2, i}^2)}.
\end{align}
If $\chi^2(P_1, P_2)$ is below a given threshold value $\chi^2_{\mathrm{thresh}}$, which is determined by the degrees of freedom\footnote{The number of degrees of freedom entering the $\chi^2$ is given by $N_{\mathrm{bins}} - 1$. Since we work with normalized distributions, they have to respect the additional constraint $\int \frac{1}{\sigma} \frac{d\sigma}{dm_{hh}} \, dm_{hh} = 1$.} $N_{\mathrm{d.o.f.}}=N_{\mathrm{bins}}-1$, and the chosen confidence level ($\mathrm{CL}$), we call the shapes $D_1$ and $D_2$ \textit{similar}, which will be denoted as
\begin{align}
    P_1 \sim P_2 \ \ \mathrm{if} \ \ \chi^2(P_1, P_2) < \chi^2_{\mathrm{thresh}}.
\end{align}
To perform an unbiased scan over the full parameter space, we first select the SM point 
\begin{align}
    P_{\mathrm{SM}} = (1, 1, 0, 0, 0, 0, 0, 0, 0, 0, 0)
\end{align}
and generate $(N_{\mathrm{sample}}-1)$ randomized points in a given parameter range, whose differential distributions only have positive values in each bin, i.e. that are physical. As a second step, we compute the symmetric $(\chi^2)_{ij}$ matrix containing the pairwise $\chi^2-$values of each combination of points. The number of independent entries of this matrix scales according to
\begin{align}
    \frac{N_{\mathrm{sample}}(N_{\mathrm{sample}}-1)}{2} \longrightarrow \frac{N_{\mathrm{sample}}^2}{2},
\end{align}
which exhibits a quadratic scaling behavior for a large number of samples. Since small numerical deviations from the SM point $P_{\mathrm{SM}}$ will also lead to small deviations from the SM distribution $D_{\mathrm{SM}}$, we first identify all points $P_i$ that are similar to the SM, i.e.~we determine all indices $i$ such that $P_i \sim P_{\mathrm{SM}}$. Next, we eliminate these points from the $(\chi^2)_{ij}$ matrix as they are well-described by the SM shape. From the remaining matrix, we search for the point with the largest number of $\chi^2-$values below the threshold value serving as the next cluster center since it approximates the largest number of shapes. If this number is the same for several points, we select the point with the smallest geometric average of the $\chi^2-$values below threshold. After again eliminating all points associated to the new cluster center, we repeat this procedure until a maximal and predetermined number of clusters $N_{\mathrm{Clus}}$ is reached or the $(\chi^2)_{ij}$ matrix is empty. Because of the quadratic scaling behavior of the number of independent entries of the matrix, the maximally feasible value for $N_{\mathrm{sample}}$ is limited. However, we can perform a validation scan with a larger number of samples $N_{\mathrm{val}}$ to check whether our selection of clusters is representative for the full parameter space of the EFT. Furthermore, this allows us to study differences between several assumptions for the bin-wise uncertainty $\Delta_i$ and the chosen number of clusters $N_{\mathrm{Clus}}$. 

\subsubsection{HEFT Results}
\begin{figure}[t] \centering
\includegraphics[width=0.9\textwidth]{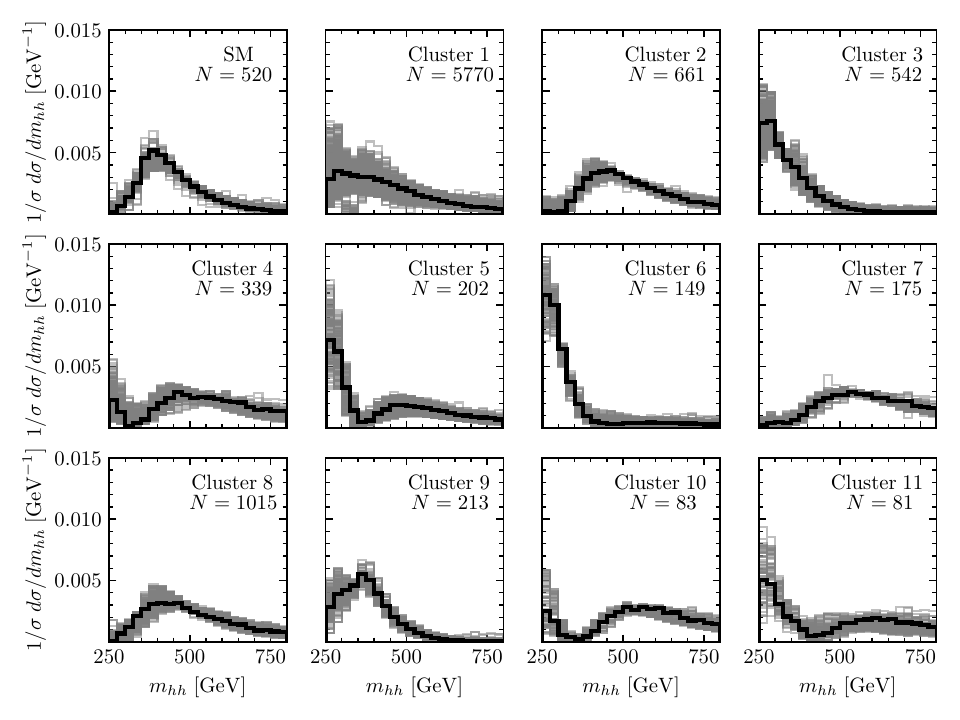}
\caption{Clusters and sample assignments obtained for the coefficients contained in $\mathcal{L}_{\kappa}$ and a number of $N_{\mathrm{samples}} = 10^4$ samples assuming the full bin-wise theory uncertainty $\Delta_{\mathrm{th}, i}$. The displayed numbers $N$ indicate how many of the generated samples can be associated to each cluster with $\chi^2 < \chi^2_{\mathrm{thresh}}$.}
\label{figure_Kappa_Delta_th}
\end{figure}

\begin{figure}[t] \centering
\includegraphics[width=0.9\textwidth]{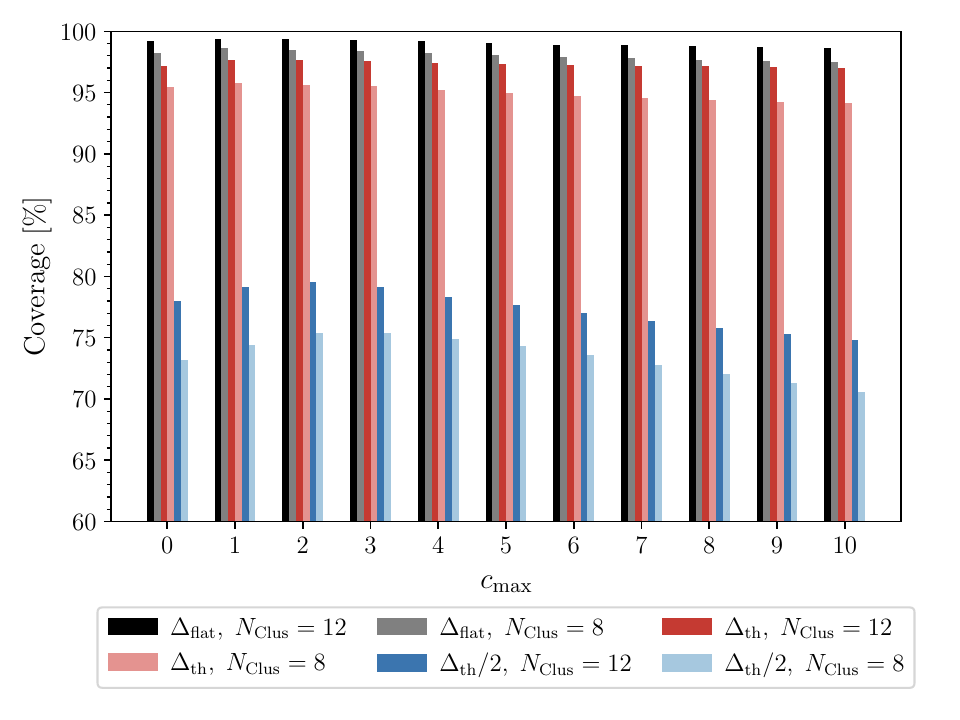}
\caption{Coverages of the $\mathcal{L}_{\kappa}$ clusters obtained for different uncertainty scenarios, HEFT parameter ranges $c_{\mathrm{max}}$, and number of clusters $N_{\mathrm{Clus}}$. For each of these validation scans, we generated $N_{\mathrm{val}} = 10^8$ samples.}
\label{figure_validation}
\end{figure}

\begin{figure}[t!] \centering
\includegraphics[width=0.7\textwidth]{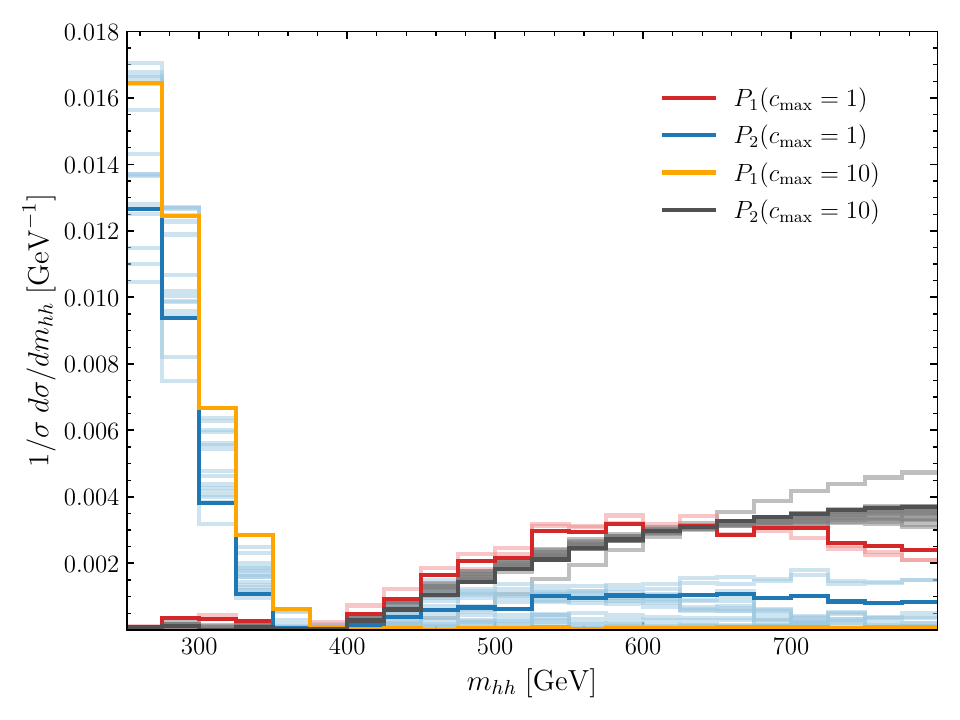}
\caption{Normalized invariant mass distributions of samples showing a large $\chi^2-$ deviation from the $\mathcal{L}_{\kappa}$ clusters assuming the full theory uncertainty and $c_{\mathrm{max}} = 1, 10$. The color choice represents the assignment of the samples to the points reported in Table \ref{table_samples_large_chi2}.}
\label{figure_samples_large_chi2}
\end{figure}

The first objective of our analysis of the HEFT invariant mass distributions is to obtain the clusters for the $\mathcal{L}_{\kappa}$ Lagrangian as was done in \cite{Carvalho:2015ttv, Capozi:2019xsi} for two different clustering procedures: in Ref. \cite{Carvalho:2015ttv}, clustering was performed using Poisson statistics applied to the number of events in each bin, with the number of clusters fixed to 12. Ref.~\cite{Capozi:2019xsi} on the other hand, employed machine-learning techniques to determine the kinematic clusters. In a first step, they used supervised learning with predetermined shapes, incorporating statistical uncertainties on the $A_i$ coefficients of the $\mathcal{L}_{\kappa}$ calculation. This was followed by an unsupervised learning step, where the number of clusters was set to either 4 or 8. In contrast, we use an estimate for the theoretical uncertainty that is composed of the PDF and $\alpha_s$, scale, and top mass scheme uncertainties. Our clustering approach continues until a number of clusters, $N_{\mathrm{Clus}}$, is reached or the number of clusters already found is exhaustive for the samples we generated, i.e. all samples can be assigned to one of the clusters with a $\chi^2 < \chi^2_{\mathrm{thresh}}$. To ensure consistency of our work with \cite{Capozi:2019xsi}, we consider the following parameter ranges\footnote{The definition of our $\mathcal{L}_{\kappa}$ Lagrangian differs slightly from the convention used in \cite{Capozi:2019xsi}. Thus, we had to rescale the numerical ranges of the coefficients $a_g$ and $b_g$ by a factor of $2$.}:
\begin{align}
    a_{\lambda^3} \in [-3, 8], \ a_t \in [0.5, 1.5], \ b_t \in [-3, 3], \ a_g \in [-0.25, 0.25], \ b_g \in [-0.25, 0.25].
\end{align}
We generate $N_{\mathrm{sample}} = 10^4$ valid samples $P_i$, that lie within this parametric region, and calculate their invariant mass distribution. To construct the $(\chi^2)_{ij}$ matrix and perform the clustering procedure, two further inputs are required: the bin-wise uncertainty $\Delta_i$ and a threshold value $\chi^2_{\mathrm{thresh}}$. We estimated the full theory uncertainty $\Delta_{\mathrm{th}, i}$ by linearly summing up a flat combined $\mathrm{PDF}$ and $\alpha_s$ uncertainty of $\Delta_{\mathrm{PDF}+\alpha_s} = 2.3 \, \%$ for the chosen PDF set \texttt{PDF4LHC15\_nlo\_100\_pdfas} \cite{Butterworth:2015oua, Dulat:2015mca}, a flat scale uncertainty $\Delta_{\mu} = 8.8 \, \%$ that corresponds to the maximal value of the NNLO uncertainties when varying $\kappa_\lambda$ as reported in Ref.~\cite{Baglio:2020wgt, Grazzini:2018bsd}, and a bin-wise top mass scheme uncertainty, which we conservatively take to be symmetric adopting the values of \cite{Bagnaschi:2023rbx}:
\begin{align}
    \Delta_{\mathrm{th}, i} = \Delta_{\mathrm{PDF}+\alpha_s} + \Delta_{\mu} + \Delta_{m_t, i}.
\end{align}
Note that $\Delta_{m_t, i}$ exhibits a strong bin dependence, varying between $2.2\,\%$ and $42.7 \, \%$ over the invariant mass range considered. The clustering threshold $\chi^2_{\mathrm{thresh}}$ is determined from $N_{\mathrm{bins}}-1 = 21$ degrees of freedom and a confidence level, which we set to $\mathrm{CL} = 99.73 \, \%$ corresponding to a significant deviation of $3\sigma$. This yields a threshold value of
\begin{align}
    \chi^2_{\mathrm{thresh}} = 43.52.
\end{align}
The resulting clustering for $N_{\mathrm{Clus}} = 12$ is shown in Fig.~\ref{figure_Kappa_Delta_th}.\footnote{We checked that using other random values, the qualitative behavior of the clusters remains the same.}
We find that $5.2 \%$ of the $10^4$ samples show a small deviation from the SM distribution $D_{\mathrm{SM}}$, while the other clusters capture several kinematic features of the shapes. These include peaks close to the invariant mass threshold of $m_{hh} = 2 m_h \sim 250 \, \mathrm{GeV}$, structures with two peaks, and flatter distributions relative to $D_{\mathrm{SM}}$. In total, the coverage of the 12 clusters is $97.62\, \%$, leaving $238$ of the $10^4$ samples unassigned under the given $\chi^2$ threshold. This observation motivates four central questions: 
\begin{enumerate}[label=(\alph*)]
\item Performing a validation scan with a number of samples $N_{\mathrm{val}} \gg N_{\mathrm{samples}}$, how does the coverage of the 12 determined clusters change for $\mathcal{L}_{\kappa}$?
\item A coverage close to $100 \, \%$ confirms that our relatively small set of kinematic clusters describes the invariant mass shapes of our full parameter space very well. How does the coverage change once we also choose non-vanishing values for the HEFT Wilson coefficients contained in $\delta\mathcal{L}$ of Eq.~\eqref{Lagrangian_Di_Higgs_HEFT} in the validation scan?
\item How does the coverage depend on the assumption for the relative uncertainty $\Delta_i$?
\item Does the coverage change significantly when choosing a smaller amount of clusters?
\end{enumerate}
To address these questions, we performed a series of validation scans with $N_{\mathrm{val}} = 10^8$ samples. We varied the HEFT coefficients over the following ranges:
\begin{align}
    b_D, d_c, a_c, b_c \in [-c_{\mathrm{max}}, c_{\mathrm{max}}], \ \ \ b_g^{(1)} \in [-c_{\mathrm{max}}, 0], \ \ \ b_g^{(2)} \in [-b_g^{(1)}, c_{\mathrm{max}}].
\end{align}
Here, the coefficient $b_g^{(1)}$ is restricted negative values, and $b_g^{(2)}$ is generated in the interval $[-b_g^{(1)}, c_{\mathrm{max}}]$ to respect the positivity bounds of Section~\ref{sec_positivity}. Furthermore, this setup ensures that the absolute value of each coefficient remains below $c_{\mathrm{max}}$. We scanned over $c_{\mathrm{max}} \in [0, 10]$ in unit steps, where $c_{\mathrm{max}} = 0$ corresponds to samples belonging to the $\mathcal{L}_{\kappa}$ Lagrangian. For each value of $c_{\mathrm{max}}$, we considered three different uncertainty scenarios: the full theory uncertainty $\Delta_{\mathrm{th}, i}$, half of the theory uncertainty $\Delta_{\mathrm{th}, i}/2$, as well as a flat uncertainty $\Delta_{\mathrm{flat}} = 28 \, \%$, which represents the flat baseline projection for the HL-LHC\footnote{The flat baseline projection assumes halved theory uncertainties and accounts for a reduction of the statistical experimental uncertainties as well as improvements of the $b$-tagging and $\tau$-reconstruction. For details consult Ref.~\cite{ATLAS:2025wdq}.} \cite{ATLAS:2025wdq}. In each of these cases, we determined the clusters of $\mathcal{L}_{\kappa}$ and performed the sample assignment for $N_{\mathrm{Clus}} = 12$ and $N_{\mathrm{Clus}} = 8$. These results for a number of $N_{\mathrm{val}} = 10^8$ samples are summarized in Fig.~\ref{figure_validation}. Across all cases, the difference in coverage between $N_{\mathrm{Clus}} = 12$ and $N_{\mathrm{Clus}} = 8$ remains below $5 \, \%$. Under the full theory and flat uncertainty scenarios, the cluster coverage ranges from $94.2\,\%$ to $99.4\,\%$, independently of the number of clusters and the specific values of the HEFT coefficients. Notably, the coverage for $\Delta_{\mathrm{th}}$ exhibits only a mild dependence on $c_{\mathrm{max}}$, which amounts to about $2\,\%$, indicating that the higher-dimensional NLO and NNLO operators absent in $\mathcal{L}_{\kappa}$ contribute only subdominantly. This behavior reflects the validity of the EFT expansion under our power counting scheme. However, when the theory uncertainty is halved, the coverage drops by $\mathcal{O}(20)\,\%$ as with a smaller relative uncertainty, the $\chi^2-$test singles out marginal differences in the differential shapes.
Also for this case, we observe only a small dependence on $c_{\mathrm{max}}$. \\ \\
To better understand the types of distributions not captured by the $\mathcal{L}_{\kappa}$ clusters, we identified 20 parameter points in the HEFT space that resulted in particularly large $\chi^2$ values under the full theory uncertainty for $c_{\mathrm{max}} = 1$ and $c_{\mathrm{max}} = 10$. For each $c_{\mathrm{max}}$ value, we group the shapes into two different categories, for each of which we display an illustrative point in Table \ref{table_samples_large_chi2}. 
\begin{table}[b!]
\centering
\renewcommand{\arraystretch}{1.3}
\setlength{\tabcolsep}{6pt}
\resizebox{\textwidth}{!}{
\begin{tabular}{l|cccccccccccc}
\hline
Sample &
$a_{\lambda^3}$ & $a_t$ & $b_t$ & $a_g$ & $b_g$ &
$b_D$ & $d_c$ & $a_c$ & $b_c$ & $b_g^{(1)}$ & $b_g^{(2)}$ &
$\chi^2$ \\\hline
$P_1(c_{\mathrm{max}} = 1)$ & 1.25 & 0.71 & 0.46 & -0.12 & -0.2 & 0.25 & 0.04 & -0.09 & 0.46 & -0.29 & 0.66 & 168.23 \\\hline
$P_2(c_{\mathrm{max}} = 1)$ & 4.67 & 1.12 & 0.13 & -0.04 & -0.23 & 0.92 & -0.61 & 0.75 & 0.71 & -0.33 & 0.72 & 136.40 \\\hline
$P_1(c_{\mathrm{max}} = 10)$ & 4.79 & 1.50 & -0.35 & 0.19 & -0.03 & -3.09 & 9.68 & -9.99 & 2.05 & -3.21 & 6.83 & 158.00 \\\hline
$P_2(c_{\mathrm{max}} = 10)$ & -0.35 & 0.83 & -0.04 & 0.15 & 0.21 & 7.02 & -2.07 & -4.29 & 3.8 & -4.59 & 9.25 & 165.77 \\\hline
\end{tabular}
}
\caption{Representative HEFT samples with $c_{\mathrm{max}} = 1$ and $c_{\mathrm{max}} = 10$ that showed a large $\chi^2-$deviation when comparing to the 12 selected clusters assuming the theory uncertainty $\Delta_{\mathrm{th}, i}$. These points exhibit shape deviations significantly exceeding the $\chi^2-$ threshold.}
\label{table_samples_large_chi2}
\end{table} \\
The invariant mass distributions corresponding to these representative points, along with the remaining high-deviation samples, are shown in Fig. \ref{figure_samples_large_chi2}.
From this analysis, we can identify two main groups of shapes not captured by the $\mathcal{L}_{\kappa}$ clustering:
\begin{enumerate}
    \item distributions with a very pronounced peak close to the Higgs pair production threshold, followed by a rapid flattening around the top pair mass threshold,
    \item distributions that remain flat up to the top pair mass threshold, then forming a plateau-like structure.
\end{enumerate}
These features are not reproduced by the invariant mass clusters obtained for the $\mathcal{L}_{\kappa}$ Lagrangian, indicating that there can be qualitatively new kinematic features in the di-Higgs invariant mass spectra. Nevertheless, such shapes remain rare in our parameter scans and do not significantly affect the clustering coverage.

\subsubsection{Comparison to SMEFT \label{sec:SMEFT}}
\begin{figure}[t] \centering
\includegraphics[width=0.75\textwidth]{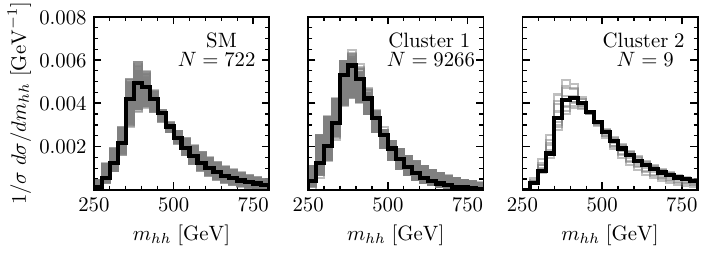}
\caption{SMEFT clusters and sample assignments obtained for $N_{\mathrm{samples}} = 10^4$ and assuming the full bin-wise theory uncertainty $\Delta_{\mathrm{th}, i}$. In the clustering process, the $\chi^2-$matrix was already exhausted by the normalized SM distribution and two additional shapes.}
\label{figure_SMEFT_Delta_th}
\end{figure}

For the purpose of a direct comparison to our HEFT results, we also apply the clustering algorithm to the SMEFT calculation of di-Higgs production. Thereby, we restrict ourselves to contributions truncated in linear order in the dimension-6 coefficients and $(16 \pi^2)^{-2}$ at the level of the squared matrix element. We parameterize the relevant part of the SMEFT Lagrangian adopting the Warsaw basis \cite{Grzadkowski:2010es} as
\begin{equation}
\begin{aligned}
\label{Lagrangian_SMEFT_Di_Higgs}
    \mathcal{L}_{\mathrm{SMEFT}} &\supset \frac{C_{H \Box}}{\Lambda^2} |H|^2 \Box |H|^2 + \frac{C_{HD}}{\Lambda^2} (H^{\dagger} D_{\mu} H)^{\ast} (H^{\dagger} D^{\mu} H) + \frac{C_H}{\Lambda^2} |H|^6 \\
    &+ \frac{C_{tH}}{\Lambda^2} |H|^2 (\bar{t}_L \tilde{H} t_R + \mathrm{h.c.}) + \frac{g_s^2}{16 \pi^2} \frac{C_{HG}}{\Lambda^2} |H|^2 G_{\mu \nu}^a G^{a \mu \nu}.
\end{aligned}
\end{equation}
Here, our parameterization of the Higgs-gluon operator reflects the assumption of a weakly-coupled and renormalizable UV completion, where this operator is loop-generated and carries an additional suppression factor of $(16 \pi^2)^{-1}$ from the UV \cite{Arzt:1994gp, Einhorn:2013kja}. In the following, we define the linear combination $C_{H, \mathrm{kin}} = C_{H \Box} - C_{HD}/4$, that enters the shift of the physical Higgs excitation $h$
\begin{align}
    h \rightarrow h + v^2 \, \frac{C_{H, \mathrm{kin}}}{\Lambda^2} \left(h + \frac{h^2}{v} \right) \, ,
\end{align}
which ensures canonical normalization of the Higgs kinetic term and removes momentum-dependent structures from the Higgs trilinear self-coupling. The contributions of the effective SMEFT operators to the di-Higgs form factors $\mathcal{M}_1$ and $\mathcal{M}_2$ can be conveniently expressed in terms of the corresponding LO SM results:
\begin{equation}
\begin{aligned}
\label{SMEFT_formfactors}
    \mathcal{M}_1  = &-\frac{g_s^2}{4 \sqrt{2} \pi^2} \frac{s(s+2 m_h^2)}{(s-m_h^2)} \frac{C_{HG}}{\Lambda^2} \\
    &+ \left[\frac{2 v^2 (s+5m_h^2)}{3 m_h^2} \frac{C_{H, \mathrm{kin}}}{\Lambda^2} - \frac{2 v^4}{m_h^2} \frac{C_H}{\Lambda^2} - \frac{s v^3}{\sqrt{2} m_h^2 m_t} \frac{C_{tH}}{\Lambda^2}\right] \, F_1^{\mathrm{SM}} \\
    &+ \frac{v^2}{m_t} \left[2 m_t \frac{C_{H, \mathrm{kin}}}{\Lambda^2} - \sqrt{2} v \frac{C_{tH}}{\Lambda^2} \right] G_1^{\mathrm{SM}}, \\
    \mathcal{M}_2 = &\frac{v^2}{m_t} \left[2 m_t \frac{C_{H, \mathrm{kin}}}{\Lambda^2} - \sqrt{2} v \frac{C_{tH}}{\Lambda^2} \right] G_2^{\mathrm{SM}}.
\end{aligned}
\end{equation}
Truncating the squared matrix element at order $\mathcal{O}(1/\Lambda^2)$, then leads to an expression for the invariant mass distribution which is linear in each of the SMEFT Wilson coefficients and reproduces the SM distribution at LO in the loop expansion in the absence of the SMEFT Wilson coefficients:
\begin{align}
    \frac{d\sigma_{\mathrm{SMEFT}}}{d m_{hh}} = A_{\mathrm{SM}}^{\mathrm{LO}} + A_{H, \mathrm{kin}} \, C_{H, \mathrm{kin}} + A_H \, C_H + A_{tH} \, C_{tH} + A_{HG} \, C_{HG},
\end{align}
where we set $\Lambda = 1 \, \mathrm{TeV}$. This parameterization of the invariant mass distributions is particularly simple as it only depends on four different coefficients, reflecting the correlations of vertices involving one and two Higgs bosons that the dimension-6 SMEFT Lagrangian (\ref{Lagrangian_SMEFT_Di_Higgs}) gives rise to. Hence, the SMEFT samples for the clustering algorithm take the form of
\begin{align}
    P_{\mathrm{SMEFT}} = (C_{H, \mathrm{kin}}, C_H, C_{tH}, C_{HG}).
\end{align}
We adopt the SMEFT parameter ranges from a recent SMEFT global analysis that reports the marginalized results including RGE effects \cite{terHoeve:2025gey} and reinterpret the previously employed bound $a_{\lambda^3} \in [-3, 8]$ as a constraint on $C_H$:
\begin{align}
    C_{H, \mathrm{kin}} \in [-4.5, 3.0], \ \ C_H \in [-16.7, 9.7], \ \ C_{tH} \in [-15.2, 3.5], \ \ C_{HG} \in [-3.6, 1.5].
\end{align}
Using these parameter ranges, we show the obtained SMEFT clusters assuming the full theory uncertainty in Fig.~\ref{figure_SMEFT_Delta_th}.
\begin{figure}[h!] \centering
\includegraphics[width=0.9\textwidth]{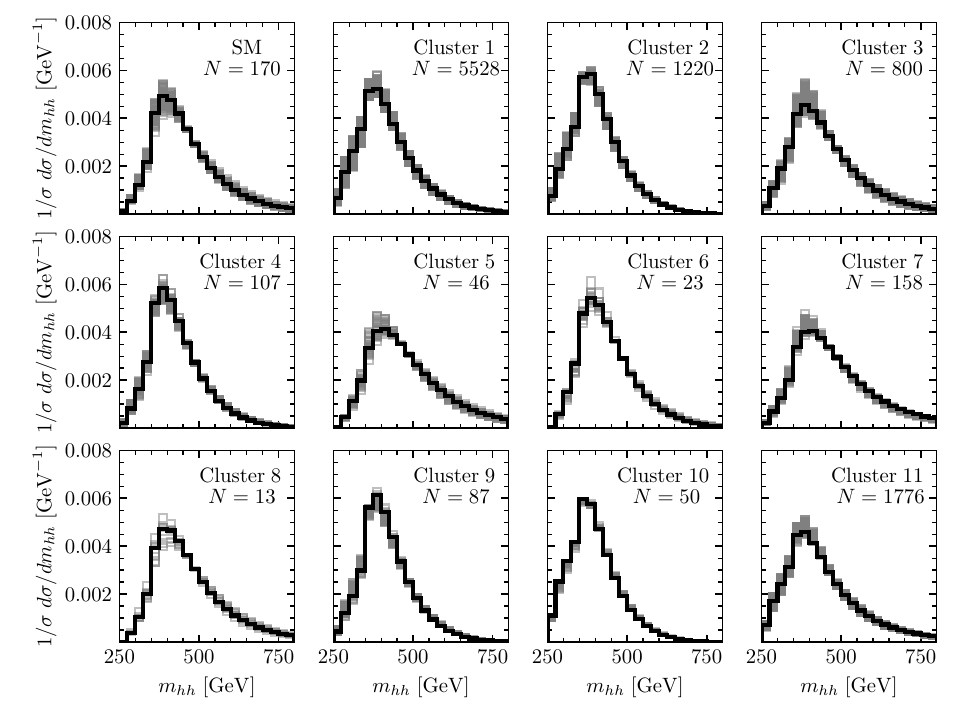}
\caption{SMEFT clusters and sample assignments obtained for $N_{\mathrm{samples}} = 10^4$ and assuming the halved bin-wise theory uncertainty $\Delta_{\mathrm{th}, i}/2$. The clustering process leads to the maximal number of $N_{\mathrm{Clus}} = 12$ clusters.}
\label{figure_SMEFT_Delta_th_half}
\end{figure}
To validate the clustering, we perform another scan with $N_{\mathrm{val}} = 10^8$ and find a coverage of $99.98 \, \%$, indicating that these four clusters cover the 4-dimensional SMEFT parameter space almost completely. When the bin-dependent theory uncertainty is halved, the full number of $N_{\mathrm{Clus}} = 12$ clusters is exhausted as shown in Fig.~\ref{figure_SMEFT_Delta_th_half}, covering $99.90 \, \%$ of the $10^4$ samples. Repeating the validation scan, yields a slightly decreased coverage of $99.78\, \%$. Given that the SMEFT prediction at this order is linear in only four parameters, with all but the $C_H$ already well constrained by single-Higgs data, even a relatively small number of clusters is sufficient to characterize the full parameter space, even under tighter uncertainty assumptions. However, this situation might change if higher-order contributions in the $1/\Lambda$ or the loop expansion were included, similar to the behavior observed in the HEFT analysis.

\subsection{Angular Observables in HEFT}
\begin{figure}[t] \centering
    \begin{minipage}{0.5\textwidth}
        \centering
    \raisebox{0.05cm}{\includegraphics[width=0.88\textwidth]{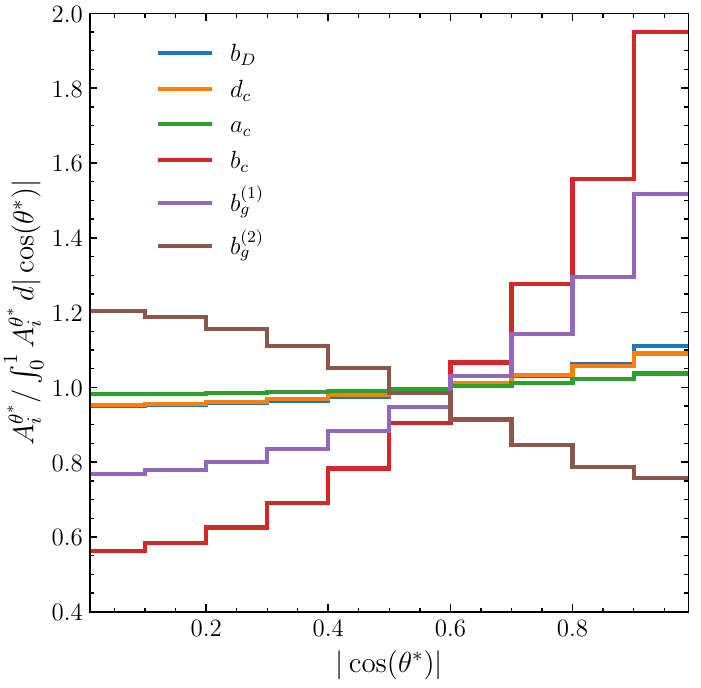}}
    \end{minipage}\hfill
    \begin{minipage}{0.5\textwidth}
        \centering
    \raisebox{0cm}{\includegraphics[width=0.91\textwidth]{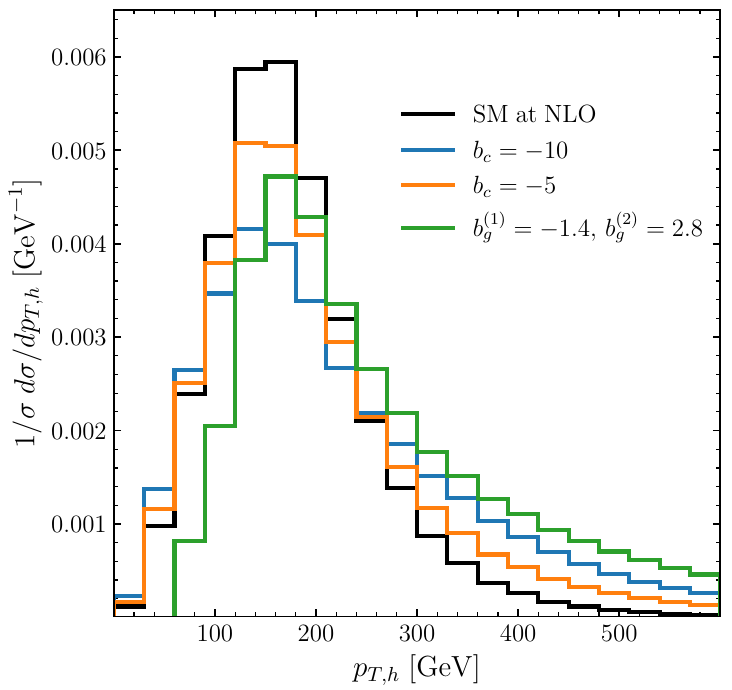}}
    \end{minipage}
\caption{Left: Comparison of the normalized bin-wise coefficients $A^{\theta^{\ast}}_i$ that enter the $|\cos{\theta^{\ast}}|-$ distribution for the HEFT coefficients $\{b_D, d_c, a_c, b_c, b_g^{(1)}, b_g^{(2)}\}$. The coefficients $b_c, b_g^{(1)},$ and $b_g^{(2)}$ show the largest activity in the scattering angle distribution. Right: Modification of the SM $p_{T, h}-$distribution at NLO QCD (black) when turning on either the $b_c$ or $b_g^{(1)}$ and $b_g^{(2)}$ coefficient. A non-zero value of $b_g^{(1)}$ automatically causes a positive non-zero value for $b_g^{(2)}$ due to the positivity bounds.}
\label{HEFT_costheta_pT}
\end{figure}
In addition to the invariant mass distributions discussed before, effective operators can also impact angular observables, in particular the scattering angle in the COM frame, $|\cos \theta^{\ast}|$, and the transverse momentum $p_{T, h}$.
\par
In the SM, the $|\cos \theta^{\ast}|$ distribution is known to be flat across the full angular range making this observable particularly sensitive to EFT modifications. Within the SMEFT framework, the results presented in Eq.~\eqref{SMEFT_formfactors} show that the only enhancement of the angular dependence can stem from the interference of the SMEFT results with the SM box form factors when squaring the matrix element. However, these effects do not give rise to  any sizable modification of the angular shape. Similar results have been found for the operators contained in the non-linear $\mathcal{L}_{\kappa}$ Lagrangian \cite{Carvalho:2015ttv, Carvalho:2016rys, Englert:2025xrc}, where only minor deviations are observed. These deviations can be enhanced by imposing additional cuts on the Higgs pair invariant mass, potentially impacting the experimental sensitivity. Consequently, the $|\cos \theta^{\ast}|$ distribution serves as a null test for the six additional operators introduced in the HEFT framework compared to the minimal non-linear $\mathcal{L}_\kappa$ setup. In the following, we study the $d\sigma/d|\cos \theta^{\ast}|$ shapes and comment on the corresponding modifications of the transverse momentum spectra.
\par
We model the angular observables using a bin-wise parameterization, analogous to Eq.~\eqref{parameterization_invariant_mass}. We introduce bin-wise coefficients $A^{\theta^{\ast}}_i$ and $A^{p_{T, h}}_i$ such that
\begin{align}
    \frac{d \sigma}{d|\cos\theta^{\ast}|} = \sum_i A^{\theta^{\ast}}_i(|\cos\theta^{\ast}|) \, c_i, \ \ \ \ \ \ \frac{d \sigma}{dp_{T, h}} = \sum_i A^{p_{T, h}}_i \, c_i.
\end{align}
In our analysis, we focus on the six HEFT coefficients $\{b_D, d_c, a_c, b_c, b_g^{(1)}, b_g^{(2)}\}$ and set the remaining HEFT to their SM values. This yields the following linear parameterizations for both angular observables $O = |\cos\theta^{\ast}|$ and $p_T$:
\begin{align}
    \frac{d \sigma}{dO} =  \frac{d\sigma_{\mathrm{SM}}^{\mathrm{NLO}}}{dO} + A^{O}_{b_D} \, b_D + A^{O}_{d_c} \, d_c + A^{O}_{a_c} \, a_c + A^{O}_{b_c} \, b_c + A^{O}_{b_g^{(1)}} \, b_g^{(1)} + A^{O}_{b_g^{(2)}} \, b_g^{(2)}.
\end{align}
We estimate the corresponding SM NLO distributions using $K-$factors as
\begin{align}
    \frac{d\sigma_{\mathrm{SM}}^{\mathrm{NLO}}}{d|\cos{\theta^{\ast}}|} = K[|\cos{\theta^{\ast}}|] \, \frac{d\sigma_{\mathrm{SM}}^{LO}}{d|\cos{\theta^{\ast}}|}, \ \ \ \ \ \
    \frac{d\sigma_{\mathrm{SM}}^{\mathrm{NLO}}}{dp_{T, h}} = K[p_{T, h}] \, \frac{d\sigma_{\mathrm{SM}}^{LO}}{dp_{T, h}}
\end{align}
with a flat $K[|\cos{\theta^{\ast}}|] = \sigma_{\mathrm{SM}}^{\mathrm{NLO}}/\sigma_{\mathrm{SM}}^{\mathrm{LO}} = 1.66$ \cite{Borowka:2016ehy} and a bin-wise $K[p_{T, h}]$ taken from \cite{Bagnaschi:2023rbx}. For the purpose of first identifying the HEFT Wilson coefficients that show a particularly large activity in the $|\cos{\theta^{\ast}}|$ observable, we examine the normalized $A^{\theta^{\ast}}$ coefficients in the left panel of Fig. \ref{HEFT_costheta_pT} using a bin width of 0.1. Notably, the coefficient $b_c$ corresponding to a chromomagnetic interaction involving two Higgs bosons, along with the Higgs-gluon contact interactions with additional derivatives described by $b_g^{(1)}$ and $b_g^{(2)}$ show a particularly large deviation from a flat behaviour. In the right panel of Fig.~\ref{HEFT_costheta_pT}, we show that these coefficients can also induce sizable modifications on the level of $p_T$ distributions. In particular, they lead to a suppression of the peak and an enhancement of the high-$p_T$ tail.  \par
\begin{figure}[h!] \centering
\includegraphics[width=0.62\textwidth]{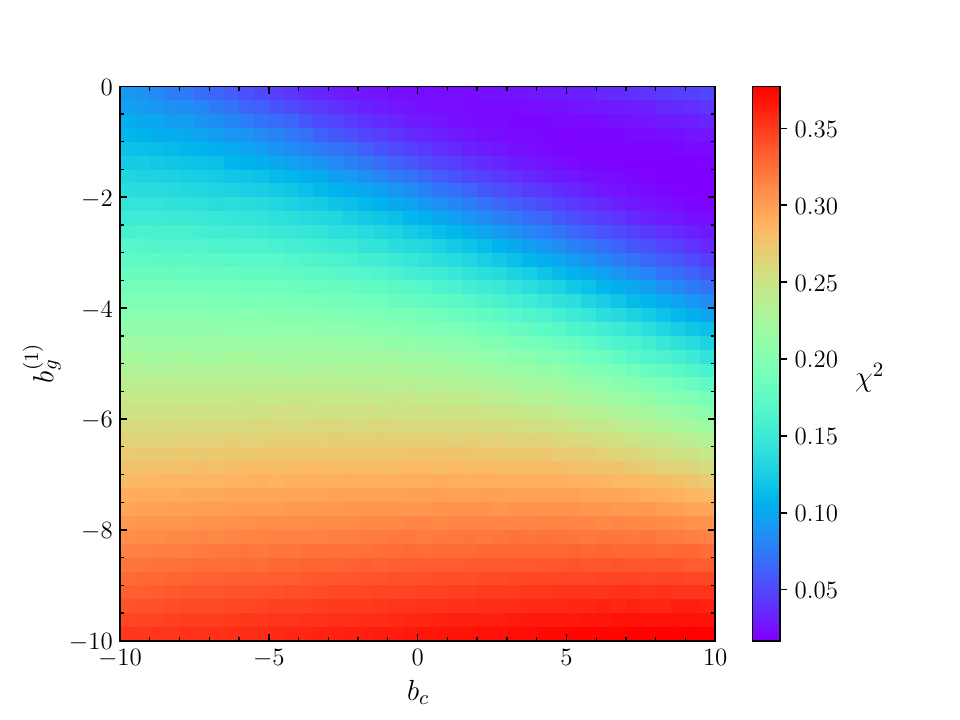}
\caption{Average $\chi^2-$values in the $b_c-b_g^{(1)}-$plane.}
\label{fig_angular_bc_bg1}
\end{figure}
\begin{figure}[h!] \centering
\includegraphics[width=0.62\textwidth]{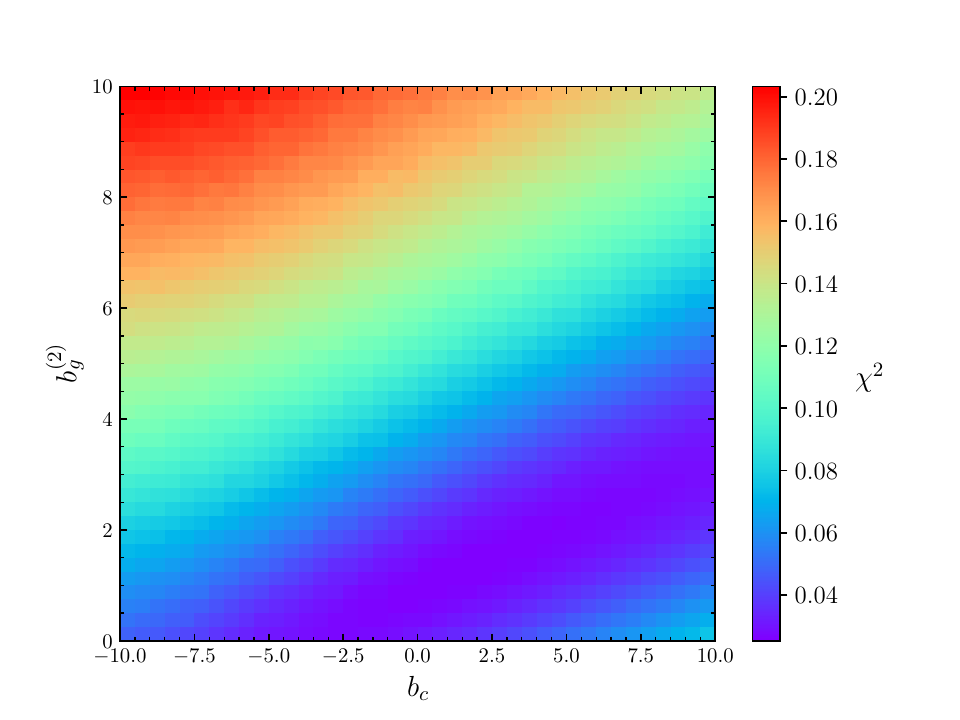}
\caption{Average $\chi^2-$values in the $b_c-b_g^{(2)}-$plane.}
\label{fig_angular_bc_bg2}
\end{figure}
\begin{figure}[t] \centering
\includegraphics[width=0.62\textwidth]{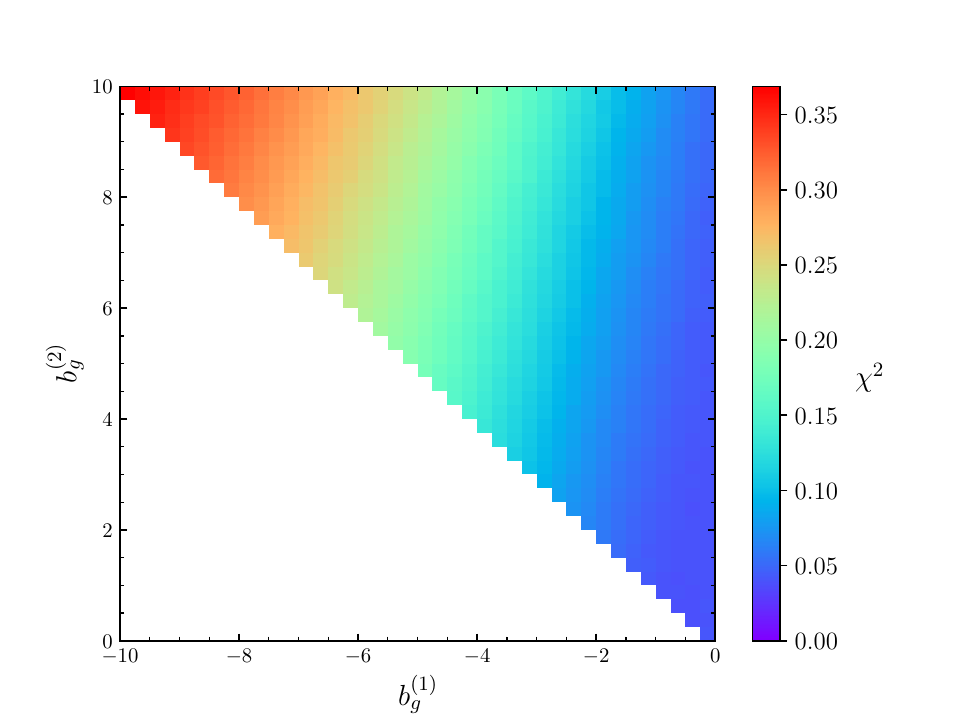}
\caption{Average $\chi^2-$values in the $b_g^{(1)}-b_g^{(2)}-$plane.}
\label{fig_angular_bg1_bg2}
\end{figure}

In order to study the scattering angle shapes in more detail, we employ a $\chi^2-$test analogous to Eq. (\ref{chi2_general_formula}). Due to the lack of a detailed uncertainty model for the scattering angle observable, unlike the invariant mass distribution, we adopt a flat uncertainty of $\Delta_{\mathrm{flat}} = 28\,\%$. To address the multi-dimensional nature of the EFT parameter space, we perform our analysis for two-dimensional projections on each of the three parameter pairs that we can form out of the three identified angularly active coefficients. We discretize the remaining two-dimensional space into $1600$ squares and generate a number of $10^5$ samples for each square also varying the other four HEFT coefficients. The corresponding $\chi^2-$ value of the square is then obtained by averaging over the $\chi^2-$values of all samples in the square. The results of this procedure are shown in Figs.~\ref{fig_angular_bc_bg1}, \ref{fig_angular_bc_bg2}, and \ref{fig_angular_bg1_bg2}. In each of the projections, an almost linear gradient is observed, which points towards $\{b_c, b_g^{(1)}, b_g^{(2)}\} \rightarrow \{-10, -10, 10\}$, showing the largest $\chi^2-$deviations. At the level of single squares, we find maximal $\chi^2$-values of order $\mathcal{O}(0.2)$, but we also obtain single samples that can achieve $\chi^2$-values of order $\mathcal{O}(10)$. Comparing to the corresponding threshold value, which for $N_{\mathrm{bins}} = 10$ and a confidence level of $3\sigma$ reads
\begin{align}
    \chi^2_{\mathrm{thresh}} = 25.26,
\end{align}
it becomes clear that these angular deviations cannot be probed without a significant improvement of the uncertainty. Under the flat uncertainty assumption, we also give an estimate of the improved value $\Delta_{\mathrm{target}}$ via
\begin{align}
    \Delta_{\mathrm{target}} < \Delta_{\mathrm{flat}} \, \sqrt{\frac{\chi^2_{\mathrm{max, analysis}}}{\chi^2_{\mathrm{thresh}}}},
\end{align}
where $\chi^2_{\mathrm{max, analysis}}$ is set to the maximal $\chi^2-$values we have identified in our analysis. Thus, on the level of single samples, we find $\Delta_{\mathrm{target}} < 17.6 \, \%$, i.e. an improvement of about $10 \, \%$. To this end, such an improvement of the uncertainty would allow us to efficiently identify deviations in the invariant mass distributions, which is why they would constitute the superior tool to discover an EFT contribution.
\newpage

\section{Conclusions}
\label{sec_conclusions}
In this work, we have computed the Higgs pair production cross section in HEFT up to truncation order $N^{\rm trunc}_{\mathrm{HEFT}}=10$, following the power counting rules of~\cite{Brivio:2025yrr}.
This requires not only the QCD-corrected diagrams of \cite{Buchalla:2018yce}, but also the inclusion of operators of chiral dimension 2 and 4 inserted into diagrams of lower loop order. The $N_{\chi}=4$ operators considered here are constrained by positivity bounds, which we derived in this work.
\par
We performed a detailed phenomenological analysis. In particular, we reviewed the cluster analysis from Refs.~\cite{Carvalho:2015ttv, Capozi:2019xsi}  currently used by the experimental collaborations to perform searches for Higgs pair production within HEFT \cite{ATLAS:2024ish, CMS:2024rgy}. In our approach, we connect the number of distinguishable kinematic distributions in $m_{hh}$ with the current and future experimental and theoretical uncertainties on this distribution. We find that, although the HEFT operators studied here for the first time can give rise to novel $m_{hh}$ distributions not yet considered in the experimental analyses, such cases occur only rarely. As a result, the coverage provided by the existing benchmark scenarios remains very good.
\par
Applied to the SMEFT at linear order, we find much less kinematic distributions in $m_{hh}$. We leave the comparison of the coverage of those benchmark scenarios with SMEFT including higher dimensional operators \cite{ggHHdim8}, quadratic terms in the cross section and/or operators appearing in higher loop countings \cite{Heinrich:2023rsd} to future work.
\par
We also looked at angular observables, in particular the $|\cos \theta^*|$ distribution which is known to be very flat both in the SM and for the LO HEFT Lagrangian. Indeed, we find that operators from the NLO and NNLO Lagrangian can modify this distribution. Mediating over the possible modifications leads us to the conclusion that the uncertainties on a measurement of the Higgs pair production $|\cos\theta^*|$ distribution need to sizably shrink to allow to distinguish such modifications of the $|\cos\theta^*|$ distribution.
For a robust identification and assessment of operators, a global approach incorporating several observables will though be essential.

\acknowledgments
We would like to thank Ken Mimasu, Alexandra Carvalho, Gudrun Heinrich and Luigi C. Bresciani for useful discussions.
The work of RG is supported by the University of Padua under the 2023 STARS Grants@UniPD Programme (Acronym and title of the project: HiggsPairs - Precise Theoretical Predictions for Higgs pair production at the LHC) and from the INFN Iniziativa Specifica APINE. The work of IB is supported by SNF through
the PRIMA grant no. 201508. This work was also partially supported by the Italian MUR Departments of Excellence grant 2023-2027 ``Quantum Frontiers''. We acknowledge support from the COMETA COST Action CA22130.\par
Our calculations made use of the \texttt{Mathematica} packages \texttt{FeynRules} \cite{Alloul:2013bka}, \texttt{FeynArts} \cite{Hahn:2000kx}, \texttt{FeynCalc} \cite{Mertig:1990an, Shtabovenko:2016sxi, Shtabovenko:2020gxv}, and \texttt{FeynHelpers} \cite{Shtabovenko:2016whf}. The shown Feynman diagrams have been created with \texttt{FeynGame} \cite{Harlander:2020cyh, Harlander:2024qbn}. \par
CloudVeneto is acknowledged for the use of computing and storage facilities.

\bibliographystyle{JHEP.bst}
\bibliography{bibliography}

\end{document}